\documentclass[preprint,apj]{emulateapj}
\usepackage{natbib}
\usepackage{graphicx}

\newlength{\colwidth}
\newlength{\fullwidth}
\newcommand{\HST}{{\em HST}}
\newcommand{\combo}{{\sc combo-17}}
\newcommand{\gems}{{\sc gems}}
\newcommand{\goods}{{\sc goods}}
\newcommand{\galfit}{{\sc galfit}}
\newcommand{\V}{F606W}
\newcommand{\z}{F850LP}

\shorttitle{GEMS quasar host galaxies at $1.8<z<2.75$}
\shortauthors{K.~Jahnke et al.}

\begin{document}

 \title{UV light from young stars in GEMS quasar host galaxies at
 $1.8<z<2.75$}

\author{K.~Jahnke\altaffilmark{1}, S.~F.~S\'anchez\altaffilmark{1},
L.~Wisotzki\altaffilmark{1,2}, M.~Barden\altaffilmark{3},
S.~V.~W.~Beckwith\altaffilmark{4,5}, E.~F.~Bell\altaffilmark{3},
A.~Borch\altaffilmark{3}, J.~A.~R.~Caldwell\altaffilmark{4},
B.~H\"au{\ss}ler\altaffilmark{3}, 
S.~Jogee\altaffilmark{4}, D.~H.~McIntosh\altaffilmark{6},
K.~Meisenheimer\altaffilmark{3}, C.~Y.~Peng\altaffilmark{7},
H.-W.~Rix\altaffilmark{3}, R.~S.~Somerville\altaffilmark{4} and
C.~Wolf\altaffilmark{8}}

\altaffiltext{1}{Astrophysikalisches Institut Potsdam, An der Sternwarte 16, 
14482 Potsdam, Germany}
\altaffiltext{2}{Universit\"at Potsdam, Am Neuen Palais 10, 14469 Potsdam, Germany}
\altaffiltext{3}{Max-Planck-Institut f\"ur Astronomie, K\"onigstuhl 17, 69117 
Heidelberg, Germany}
\altaffiltext{4}{Space Telescope Science Institute, 3700 San Martin Drive, 
Baltimore MD, 21218, USA}
\altaffiltext{5}{Johns Hopkins University, 3700 San Martin Drive, 
Baltimore MD, 21218, USA}
\altaffiltext{6}{Department of Astronomy, University of Massachusetts, 
710 North Pleasant Street, Amherst, MA 01003, USA}
\altaffiltext{7}{Steward Observatory, University of Arizona, 933 N.\ Cherry
Ave., Tucson AZ, 85721, USA}
\altaffiltext{8}{Department of Physics, Denys Wilkinson Bldg., University
of Oxford, Keble Road, Oxford, OX1 3RH, UK}
\email{kjahnke@aip.de}

\begin{abstract}
We have performed \HST\ imaging of a sample of 23 high-redshift
($1.8<z<2.75$) Active Galactic Nuclei, drawn from the \combo\
survey. The sample contains moderately luminous quasars ($M_B \sim
-23$). The data are part of the \gems\ imaging survey that provides
high resolution optical images obtained with the Advanced Camera for
Surveys in two bands (\V\ and \z), sampling the rest-frame UV flux of
the targets. To deblend the AGN images into nuclear and resolved (host
galaxy) components we use a PSF subtraction technique that is strictly
conservative with respect to the flux of the host galaxy. We resolve
the host galaxies in both filter bands in 9 of the 23 AGN, whereas the
remaining 14 objects are considered non-detections, with upper limits
of less than 5~\% of the nuclear flux. However, when we coadd the
unresolved AGN images into a single high signal-to-noise composite
image we find again an unambiguously resolved host galaxy. The
recovered host galaxies have apparent magnitudes of
$23.0<\mathrm{\V}<26.0$ and $22.5<\mathrm{\z}<24.5$ with rest-frame UV
colours in the range
$-0.2<(\mathrm{\V}-\mathrm{\z})_\mathrm{obs}<2.3$. The rest-frame
absolute magnitudes at 200~nm are
$-20.0<M_{200~\mathrm{nm}}<-22.2$. The photometric properties of the
composite host are consistent with the individual resolved host
galaxies. We find that the UV colors of all host galaxies are
substantially bluer than expected from an old population of stars with
formation redshift $z\le5$, independent of the assumed
metallicities. These UV colours and luminosities range up to the
values found for Lyman-break galaxies (LBGs) at $z=3$. Our results
suggest either a recent starburst, of e.g.\ a few per cent of the
total stellar mass and 100~Myrs before observation, with mass-fraction
and age strongly degenerate, or the possibility that the detected UV
emission may be due to young stars forming continuously. For the
latter case we estimate star formation rates of typically
$\sim$$6\,\mathrm{M}_\odot\;\mathrm{yr}^{-1}$ (uncorrected for
internal dust attenuation), which again lies in the range of rates
implied from the UV flux of LBGs. Our results agree with the recent
discovery of enhanced blue stellar light in AGN hosts at lower
redshifts.
\end{abstract}


\section{Introduction}\label{sec:intro}
Around redshifts of $z\sim 2$--3, luminous quasars were orders of
magnitude more numerous than today. Although the physics of how active
galactic nuclei evolve is still not understood, several links between
galaxy and quasar evolution have emerged over recent years.  The
observational confirmation of supermassive black holes in the nuclei
of all galaxies with a substantial bulge component
\citep[e.g.][]{gebh00} makes every such galaxy a potential AGN
host. The strong evolution of the AGN space density could therefore be
related to the availability of accretion fuel in the host galaxies, or
to the frequency of AGN triggering events.

Gravitational interaction and major or minor merging of galaxies have
long been suggested as important factors in driving nuclear
activity in galaxies. Confirming any of these as the dominant process
has proved difficult, mainly because the morphological characteristics
found for relatively nearby AGN host galaxies are so
diverse. Furthermore, the properties of the hosts in the `heyday' of
quasars ($z \ga 2$) are still elusive, a consequence of the
substantial observational difficulties. The contrast between the
bright nuclear point sources and the surrounding galaxy increases
dramatically beyond $z\sim 1$ as a result from both surface brightness
dimming and waveband shifts towards the rest frame UV.

The last years have seen numerous attempts to resolve the host
galaxies of high-redshift quasars. Owing to the observational
challenges of detecting distant host galaxies the observational effort
for each object is large, and the observed samples have consequently
been very small, of the order of $\la 5$ per target group. While
radio-loud quasars appear to be very extended and have been resolved
out to $z \sim 4$
\citep[e.g.,][]{lehn92,carb98,hutc99,kuku01,hutc03,sanc03}, this is
not the case for the large majority of radio-quiet quasars.

At high redshifts two constraints dominate observational studies of
host galaxies: On one side, very good seeing conditions are required
to maximize the spatial contrast of the compact nuclear source
compared to the extended host galaxy. On the other, large telescope
apertures are preferrential to trace faint quasar hosts to as far away
from the nucleus as possible. Thus significant progress had to wait
for 8m-class telescopes at very good sites with active optics systems
-- with a very high light collecting power but atmospheric seeing
limitations -- and for the \HST\ and its high space-based sensitivity,
combined with unprecedented spatial resolution, but limited size that
might miss light from faint outer structures of the hosts. Some host
galaxies of radio-quiet quasars at $z\simeq 2$ have now been resolved
both in the near infrared \citep{aret98,kuku01,ridg01,falo04} and in
the optical domains \citep{hutc02}, showing these objects to be
moderately luminous, corresponding to present-day $L^\star$ or
slightly brighter.

However, host galaxy colours have been unavailable, precluding
estimates of the mass-to-light ratio ($M/L$).  Thus, without colours
the observed luminosities, and their evolution with redshift, cannot
be mapped to the mass evolution if young stars contribute a major
fraction of AGN host's light. This is important as several
high-luminosity quasars at $z \ga 2$ appear to be located in very
UV-luminous host galaxies \citep{lehn92,aret98,hutc02}. Also, at low
redshifts there is a link between nuclear activity and enhanced global
star formation in the host galaxies. \citet{kauf03} reported that SDSS
spectra of local Seyfert~2 galaxies show a significant contribution
from young stellar populations, and that this trend is strongly
correlated with nuclear luminosity.  In a multicolour study of low-$z$
QSO hosts \citep{jahn04a} as well as at intermediate redshifts
\citep[see below]{sanc04a} we found that hosts of elliptical
morphology can be significantly bluer than the bulk of inactive
ellipticals. These results indicate that in the recent past the star
formation activity in galaxies hosting an AGN may be different from
normal galaxies. The details are far from understood. Clearly more
information is required to investigate the relation of starformation
and AGN activity, their common cause or causal order and the evolution
of these properties with redshift.

The new generation of wide-field imaging mosaics obtained with the
Hubble Space Telescope (\HST), especially in the conjunction with deep
AGN surveys, has opened a new observational avenue towards AGN host
galaxy studies. Here we present first results on AGN within the
\gems\ project \citep{rix04}, the largest \HST\ colour mosaic to date.
In the present paper we investigate the presence of rest-frame
ultraviolet light in a substantial sample of $z>1.8$ AGN, all with
nuclear luminosities near $M_B = -23$. In a companion paper
\citep{sanc04a} we study rest-frame colours and morphological
properties of a sample of intermediate-redshift ($z \la 1$) AGN.

The paper is organised as follows. We first describe the sample
selection and properties together with a summary of the observational
data (Sect.~\ref{sec:data}). We then comment on the decomposition of
the nuclear and galaxy contribution, including a brief summary of the
extensive simulations that we use to estimate measurement errors
(Sect.~\ref{sec:analysis}).  In Sect.~\ref{sec:results} we present the
measured host galaxy magnitudes and describe our treatment of
non-detections.  We move on to discuss the results in
Sect.~\ref{sec:discussion}, followed by our conclusions in
Sect.~\ref{sec:conclusions}.  We use
$H_0=70$\,km\,s$^{-1}$\,Mpc$^{-1}$, $\Omega_m=0.3$ and $\Omega_\Lambda
= 0.7$ throughout this paper. All quoted magnitudes are zeropointed to
the AB system with ZP$_\mathrm{F606W}=26.493$ and
ZP$_\mathrm{F850LP}=24.843$.

\section{AGN in the GEMS survey}\label{sec:data}

\subsection{Overall survey properties}

\gems, Galaxy Evolution from Morphologies and SEDs \citep{rix04} is a
large imaging survey in two bands (F606W and F850LP) with the Advanced
Camera for Surveys (ACS) aboard \HST. Centered on the Chandra Deep
Field South (CDFS), it covers an area of $\sim 28'\times28'$ (78 ACS
fields).  Each ACS field was integrated for $3\times 12$--13\,min
exposures per filter (one orbit), dithered by $3''$ between exposures.
The individual images were then combined, corrected for the ACS
geometric distortion, and at the same time rebinned to a finer pixel
grid of $0\farcs03$, achieving approximate Nyquist-sampling of the
PSF.  The image combination also removed artefacts such as cosmic ray
hits and satellite trails. Resulting point source limiting magnitudes
are $m_\mathrm{AB}(\mathrm{F606W})=28.3$ ($5\sigma$) and
$m_\mathrm{AB}(\mathrm{F850LP})=27.1$ ($5\sigma$).  In its central
$\sim$1/5, \gems\ incorporates the epoch~1 data from the \goods\
\citep{giav03} project, that are similarly deep as the other \gems\
fields.  Further details of the data reduction procedure will be given
in a forthcoming paper (Caldwell et al.\ 2004, in prep.).

The area covered by \gems\ coincides with one of four fields covered
by the \combo\ survey \citep{wolf04}, which produced a low-resolution
spectrophotometric data base (based on photometry in 17 filters) for
about 10\,000 galaxies and 60 type~1 AGN brighter than $R\la 24$ (Vega
zeropoint) in the CDFS area \citep{wolf03b,wolf04}. The large number
of filters permitted simultaneous assignment of accurate SEDs and
redshifts for both galaxies and type~1 AGN. Galaxies and AGN are
classified by matching an SED template library to the set of 17
photometric points. The AGN SED is composed of a range of continuum
spectra with added broad emission lines \citep[all details are given
in][]{wolf04}. Type~2 AGN as well as very low luminosity AGN are
invariably classified as galaxies.  \combo\ photometric redshifts are
very reliable, with an rms scatter of $\sigma_z/(1+z) \simeq 0.02$ for
galaxies (at $z < 1.2$) and $\sigma_z/(1+z) \simeq 0.03$ for AGN at
all redshifts.  In this paper we address specific \combo\ sources just
by their running identifiers; the full \combo\ list of classifications
in the CDFS will be made available in the future (Wolf et al.\ 2004,
in prep.).

\begin{figure}
\includegraphics[clip,angle=0,width=8.5cm]{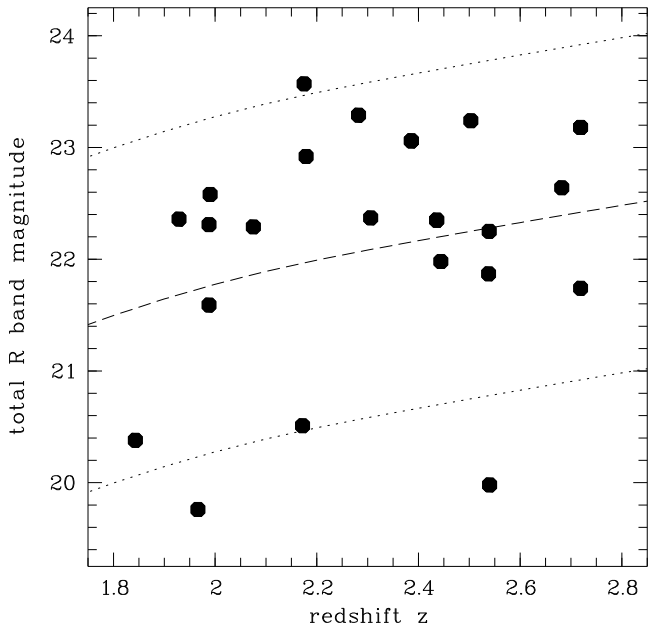}
\caption{\label{z_R} 
\combo\ redshifts and $R$ magnitudes of the AGN in the sample. 
The dashed line marks the expected $R$ band magnitude for an AGN
of $M_B=-23$ at the corresponding redshifts, assuming a typical
quasar spectrum \citep{wolf03}, the dotted lines correspond to
$M_B=-21.5$ and $M_B=-24.5$ for reference.
}
\end{figure}

\subsection{The AGN sample}\label{sec:gems}

The \combo -selected AGN in the \gems\ field range over redshifts from
$z\simeq 0.5$ up to $z\simeq 4$. In this study we investigate the
high-$z$ part of the AGN distribution.  Our sample contains all AGN
brighter than $R = 24$ in the redshift range $1.8 < z < 2.75$ that
show a meaningful counterpart in the \gems\ images. This excludes
three objects with a primary classification as AGN by \combo\ that
were apparently low redshift ($z\sim0.1$) emission line galaxies.

At these redshifts, all light detected in the \gems\ bands will
originate in the rest frame ultraviolet.  In fact, the lower redshift
limit has been imposed to ensure that the long wavelength cutoff of
the \z\ filter is still located below the Balmer jump for all
objects. The upper redshift boundary, on the other hand, was set to
avoid contamination from possible extended Ly$\alpha$ emission in the
\V\ filter. The resulting sample contains 23 AGNs, three of which are
positioned in the overlap region of two tiles so two separate images
exist for these. One object (\combo\ 05696) shows inconsistent
photometry between the two tiles due to variability over the 111 days
between the two integrations. Another object (\combo\ 19965) was
classified in \combo\ with a redshift of $z=0.634$ that had to be
revised following spectroscopy observations (G.\ Worseck, private
communication) and now entered this sample with
$z=1.90$. Table~\ref{tab:sample} gives an overview over the sample
properties, and Figure~\ref{z_R} shows the distribution in $R$ and
$z$. The average absolute magnitudes for these objects -- which we
call intermediately luminous quasars -- place them close to the
canonical division of $M_B \simeq -23$ between Seyfert galaxies
(low-luminosity) and QSOs (high-luminosity).

Several X-ray sources in the CDFS have already been studied by \HST\ with
the WFPC2 camera \citep{schr01,koek02,grog03}, most of them faint AGN,
but their sample is completely disjoint from the \combo\ AGN selection 
for this redshift range. One object falls into our redshift range 
but has $R>24$, outside our selection limits.

\begin{table*}
\begin{center}
\caption{Objects of the sample.\label{tab:sample}}
\begin{tabular}{cccccccc}
\tableline
\tableline
ID\tablenotemark{a}&
Tile\tablenotemark{b}&
RA (2000)&
DEC (2000)&
$z$\tablenotemark{c}&
$R$ (Vega)\tablenotemark{d}&
\V\tablenotemark{e} &
\z\tablenotemark{e} \\
\tableline
12325 &11& 03\,33\,01.7& --27\,58\,19& 1.843& 20.38& 20.13& 19.65\\
19965 &23& 03\,31\,45.2& --27\,54\,36& 1.90 & 19.96& 20.59& 20.29\\
30792 &82& 03\,32\,43.3& --27\,49\,14& 1.929& 22.36& 21.60& 22.06\\
02006 &04& 03\,32\,32.0& --28\,03\,10& 1.966& 19.76& 19.59& 18.98\\
04809 &08& 03\,31\,36.3& --28\,01\,50& 1.988& 22.31& 21.18& 20.91\\
06817 &09& 03\,31\,27.8& --28\,00\,51& 1.988& 21.59& 21.61& 20.86\\
18324 &19& 03\,33\,00.9& --27\,55\,22& 1.990& 22.58& 22.00& 21.33\\
05498 &01& 03\,33\,16.1& --28\,01\,31& 2.075& 22.29& 22.91& 22.28\\
11941 &10& 03\,33\,26.3& --27\,58\,30& 2.172& 20.51& 20.80& 20.40\\
62127 &62& 03\,31\,36.7& --27\,34\,46& 2.175& 23.57& 24.91& 24.37\\
51835 &55& 03\,31\,40.1& --27\,39\,17& 2.179& 22.92& 23.01& 22.41\\
00784 &05& 03\,32\,27.1& --28\,03\,36& 2.282& 23.29& 23.28& 22.80\\
36120 &39& 03\,31\,49.4& --27\,46\,34& 2.306& 22.37& 22.70& 22.23\\
05696 &02& 03\,33\,21.8& --28\,01\,21& 2.386& 23.06& 22.73& 22.32\\
05696 &03&    $''$     & $''$        & $''$ & $''$ & 23.14& 22.68\\
07671 &07& 03\,31\,51.8& --28\,00\,26& 2.436& 22.35& 22.35& 22.24\\
07671 &15&    $''$     & $''$        & $''$ & $''$ & 22.36& 22.22\\
06735 &02& 03\,33\,06.3& --28\,00\,56& 2.444& 21.98& 22.14& 21.88\\
01387 &08& 03\,31\,44.0& --28\,03\,20& 2.503& 23.24& 24.05& 23.17\\
33644 &31& 03\,32\,59.9& --27\,47\,48& 2.538& 21.87& 21.28& 21.17\\
11922 &11& 03\,33\,09.1& --27\,58\,27& 2.539& 22.25& 22.65& 21.93\\
16621 &19& 03\,33\,09.7& --27\,56\,14& 2.540& 19.98& 20.41& 20.06\\
15396 &21& 03\,32\,16.2& --27\,56\,44& 2.682& 22.64& 22.69& 22.41\\
33630 &33& 03\,31\,40.1& --27\,47\,46& 2.719& 21.74& 22.21& 21.98\\
42882 &45& 03\,32\,01.6& --27\,43\,28& 2.719& 23.18& 23.89& 23.48\\
42882 &95& $''$        & $''$        & $''$ & $''$ & 24.04& 23.54\\
\tableline
\end{tabular}
\end{center}
\tablenotetext{a}{ID from the \combo\ catalogue}
\tablenotetext{b}{\gems\ tile number (1--63 \gems, 80--95 \goods\
region)} 
\tablenotetext{c}{Photometric redshift from \combo; for \combo\ 19965
a revised redshift from an ongoing objective prism survey was used}
\tablenotetext{d}{$R$-band magnitude (Vega zeropoint) from \combo}
\tablenotetext{e}{Total \V- and \z-band magnitudes (AB zeropoint) as
measured with the ACS. Photometric errors range from 0.07--0.10 for
both filter. Three objects are imaged on two different tiles. Due to
nuclear variability this can result in different \V\ and \z-band
fluxes.}
\end{table*}

\section{Data analysis}\label{sec:analysis}
\subsection{Background and variances}\label{sec:bg}

Even though space based, ACS shows a non-negligible background from
stray light. In the reduction process already a global, outlier
clipped median background was subtracted (Caldwell et al.\ 2004, in
prep.). As the deblending of nuclear and host galaxy component with
two-dimensional modelling is sensitive to background sources, we
applied an extra procedure to remove net residuals in the local
background. This included an iterative masking of all objects in the
field and the determination of the local background from the
object-free regions.  For each square of $200\times200$ pixels an
average from the unmasked pixels was computed, with a subsequent
bilinear interpolation between these values to yield a background
estimate for the whole field.  After background adjustment, small
subimages of $128\times128$ pixels were extracted around all AGNs,
corresponding to a field of view of $3\farcs84\times3\farcs84$. This
field size contains $\ge 99$~\% of the ACS PSF flux. Since the AGN are
not strongly resolved this fraction also applies to the AGN flux.

The data reduction procedure kept record of the individual pixel
weights throughout the process of reduction and combination.  This
information was then used in combination with the shot noise derived
from pixel count rates to construct variance images which were later
used in the error budget calculations.

\subsection{PSF estimation}\label{sec:psf}

\begin{figure*}
\epsscale{0.7}
\plotone{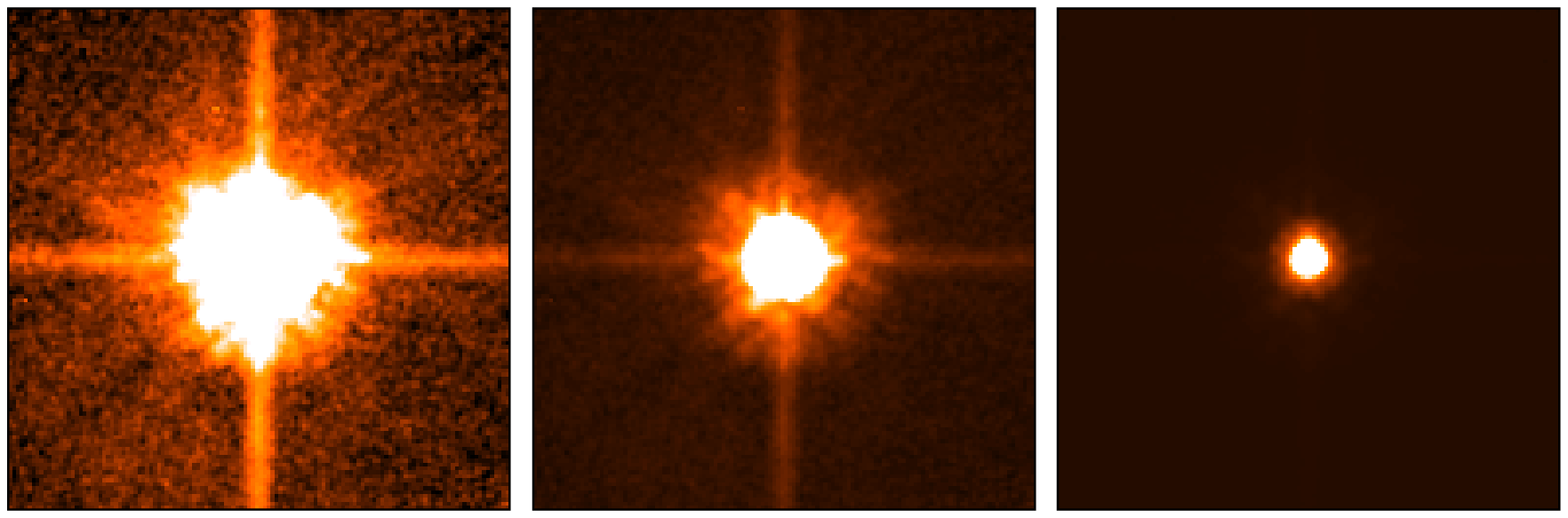}\\[2mm]
\plotone{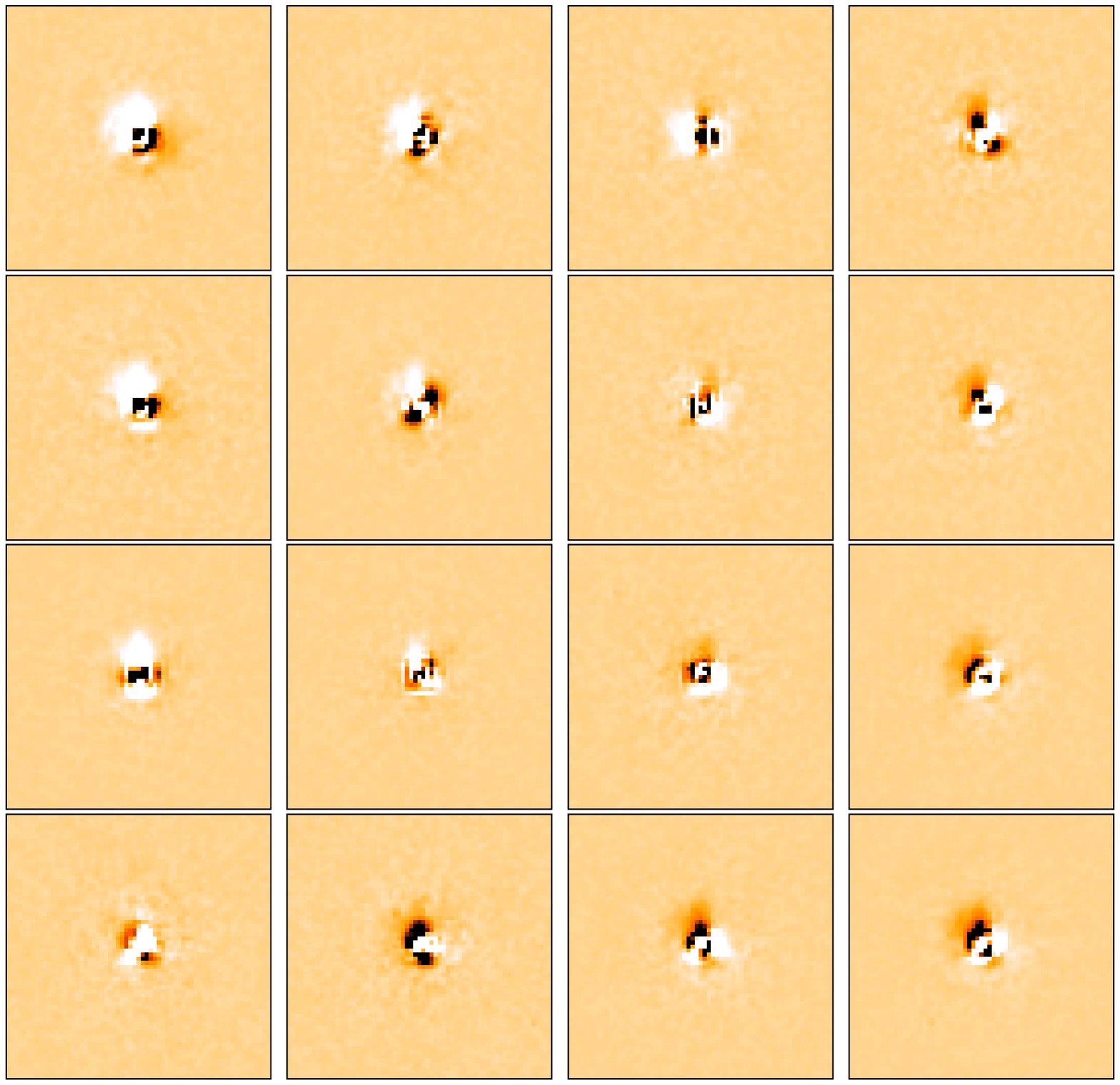}
\caption{\small \label{fig:psf} 
Top: PSF in the \V-band, shown with three display levels to enhance
structures (size is 3\farcs84$\times$3\farcs84). Bottom: Subtraction
of a total mean PSF from the PSFs in each of 4$\times$4 subtile
regions to show the spatial variations of the PSF in the inner region
(size is 1\farcs92$\times$1\farcs92).
}
\end{figure*}

Compared to ground based telescopes, the point spread function (PSF)
of ACS is very stable in time. However, coma, astigmatism and defocus
from surface height variations of the two CCDs lead to variations over
the field of view that need to be taken into account. The variations
are much weaker than for the WFPC2 camera but remain non-negligible.
Also the `breathing' of \HST\ changes the focal length which leads
also to a small time variability in the PSF. In Figure~\ref{fig:psf}
we show the mean ACS PSF compiled from $\sim$500 stars in the
extensive \gems\ area and the variation of the PSF over the FOV.

The variations were also investigated by \citet{kris03b} from the
crowded field of 47~Tucanae. Such an analysis was not possible for
individual \gems\ fields due to the small number of stars ($\sim$10)
per field. Thus a simultaneous characterisation of the spatial and
temporal variations was not possible. However, since the spatial
variations dominate (further details of our investigations of the ACS
PSF variations will be given in a dedicated technical paper, Jahnke et
al.\ 2004, in prep.) we used the large number of unsaturated stars in
the \gems\ area to construct an empirical PSF individually for each
object. At a given position we combined the nearest $\sim$35
undisturbed stars to create a position-specific PSF estimate. In this
way we average over time, but only stars from a radius $\la40\arcsec$
were used, and PSF shape errors due to {\em spatial} variation were
minimized, while a very high S/N was achieved for each of these PSF
estimates. Finally, the subimage of each AGN and its connected PSF
were registered to a common centroid.

From our PSF analysis we found that while coherent large-scale
variations were essentially absent within each stack of $\sim$35 PSF
stars, there was still considerable mismatch between the individual
stars, in particular in the central pixel regions. As such mismatched
pixels could be spuriously assigned to a host galaxy, we took the
variations within each PSF stack to derive rms frames describing an
inherent PSF uncertainty; these were then also included in the
variance images, artificially reducing the weight in the inner pixel
regions. PSF rms errors per pixel range from up to 30\% in individual
pixels inside 1~FWHM to 5--15~\% inside $0\farcs2$, and to
generally below 5\% outside.

\subsection{Peak scaled PSF subtraction}\label{sec:psfsub}

For luminous AGN at high redshifts, separating the galaxy image from
the nuclear point source is a daunting task. Even with a very good
knowledge of the PSF, the problem is still that the relative scalings
of galaxy and AGN are not known {\em a priori}. In fact, one cannot
even be certain that the host galaxy is detectable at all and not
swamped by the central point source. We therefore started out with the
well-established technique of simple PSF subtraction.

In each case, the PSF was scaled to the central flux of the AGN,
integrated inside a circular aperture of 4 pixel (0\farcs12) diameter
centered on the nucleus. This radius encircles approximately 34\% of
the total energy of a point source. A smaller radius (e.g.\ 1 pixel)
would become too sensitive to shot noise and PSF mismatch, while
larger radii contain a significantly higher fraction of the total flux
(50\%/60\%/70\% at 3/4/5 pixel radius) and thus would make a detection
of any host galaxy component successively harder.

This procedure somewhat oversubtracts the nuclear component by an
amount corresponding to the underlying host galaxy contribution inside
the encircled region. However, this method yields strictly
conservative estimates the host galaxy flux, i.e.\ always
unserestimating it. We used extensive simulations to determine
correction factors for this oversubtraction (see
Section~\ref{sec:errors}). The peak-scaled PSF subtraction method has
the advantage of being independent of any assumption about the host
galaxy morphology.

\subsection{Two-dimensional deblending}\label{sec:galfit}

As a second method we employed the modelling package \galfit\ in
Version 1.7a \citep{peng02} that allows the simultaneous fitting of
several two-dimensional components to an image, convolved with a given
PSF. We describe our application of \galfit\ to quasar images in
\citet{sanc04a} in detail.

In total we ran \galfit\ in three configurations, always fitting two
components, which were the point-source nucleus and either an
exponential disk \citep{free70}, a $r^{1/4}$ de~Vaucouleurs spheroid
\citep{deva48}, or a \citet{serc68} model with the S\'ersic index as a
free parameter. However, simulations (Sect.~\ref{sec:errors}) showed
that with the present data, the \galfit\ version used\footnote{The
\galfit\ version 1.7a was recentering the PSF to given coordinates
using a convolution with a narrow gaussian, not by shifting the PSF by
means of rebinning to a new position. The latter is better for the
application to AGN decomposition and is now incorporated in later
versions of \galfit.}  was operating near its limit, due to the very
high contrast between nuclei and host galaxies. In the context of this
paper we thus used \galfit\ only as a cross-check on the peak
subtraction method.

\subsection{Detection sensitivity}\label{sec:detection}

To determine the limits for detecting host galaxies we constructed a
sample of 200 randomly selected unsaturated stars, 100 in each
observed band, to mimic unresolved quasars. This way we could
investigate how our nucleus-removal techniques responded to a
undetectable host galaxy, and we could set limits on the size and
shape of expected residuals, thus lower flux limits for detectable
host galaxies. For these stars the PSFs were created in exactly the
same way as for the AGNs. The object itself was always excluded in the
PSF production thus each test star and its PSF were fully independent.

This set of simulated `naked quasars' showed that in 88\% (97\%) of
all cases, any residuals -- which could be taken as spurious (g)host
galaxy detections -- had fluxes of less than 5\% (10\%) of the total
object flux.  From this we adopted the condition that a real detection
should show a residual flux after peak subtraction of at least 5\% of
the total flux, corresponding to a maximum nucleus-to-host ratio of
20. Because of the systematic oversubtraction inherent in the
procedure corrections for the flux have to be applied (see next
section).

The final decision if a host galaxy is resolved is based on this
criterion. In addition we visually inspected whether the detected flux
indeed came from a host galaxy or whether other, unmasked structures
were present, using the peak subtracted images and radial profile. If
this could be ruled out we classified a host galaxy as detected.

\begin{figure*}
\includegraphics[clip,angle=0,width=8cm]{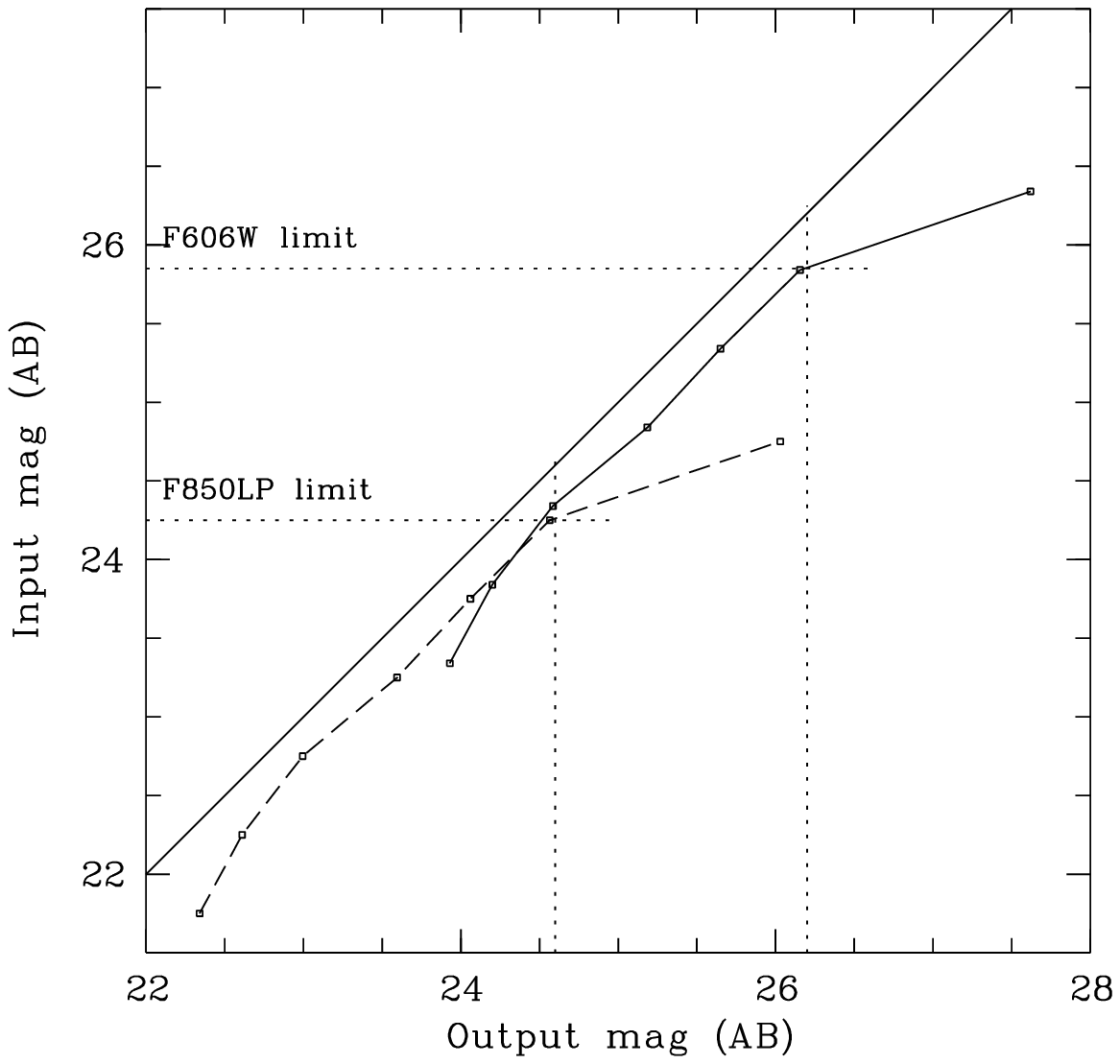}
\hfill
\includegraphics[clip,angle=0,width=8cm]{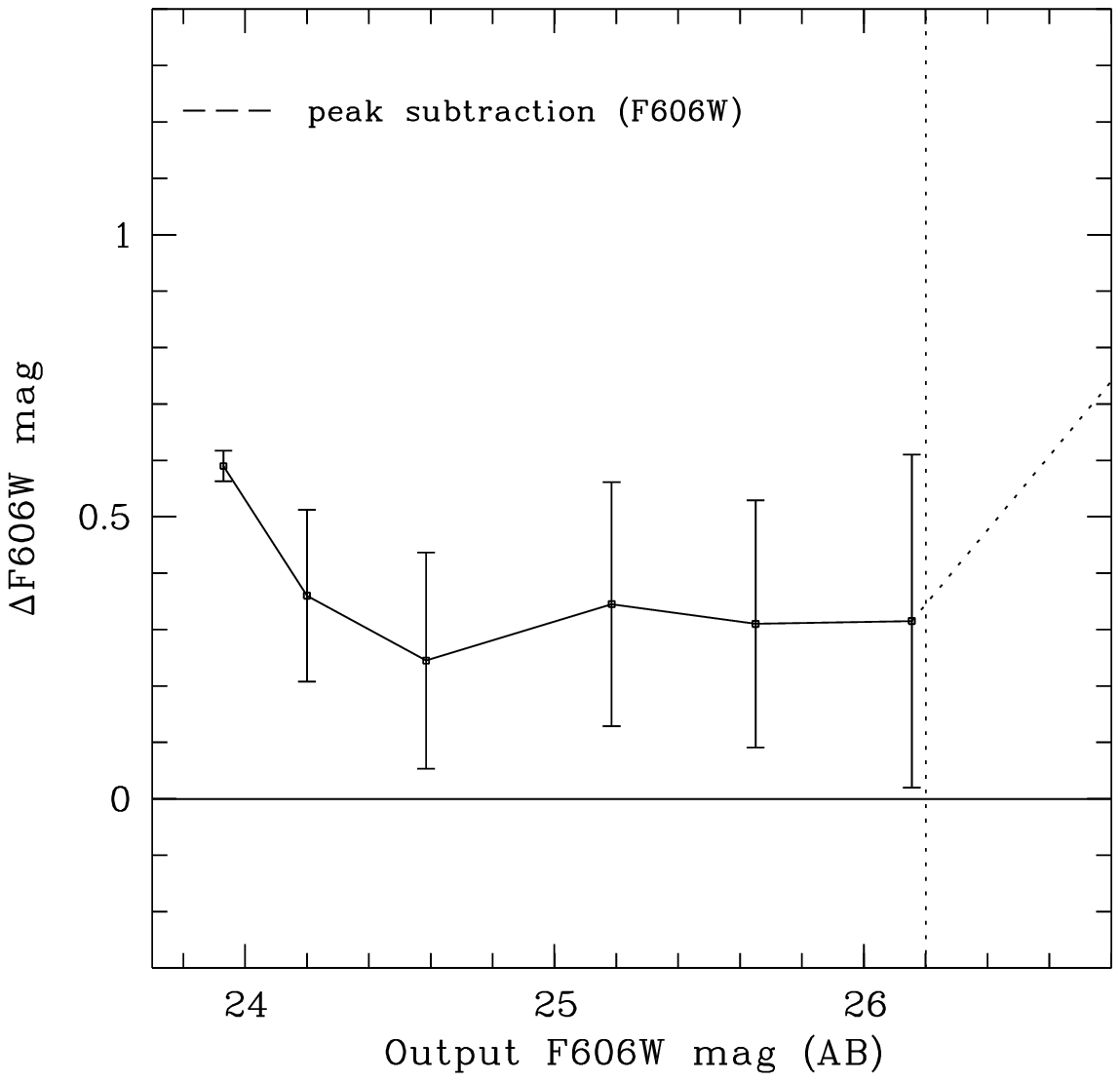}
\caption{\label{fig:corrs_and_errors_z}
Simulation results on systematic offsets and errors in recovered host
galaxy magnitude. Left: Recovered host magnitude as function of input
magnitude for a range of morphologies, differing in scales and
nucleus-to-host flux ratios matching our sample, for the two
bands. The adopted reliability regions end at the dotted straight
lines (see text). The \V- and \z-bands are simply shifted with respect
to each other by their zero-point offset
($\mathrm{ZP}_\mathrm{\V}-\mathrm{ZP}_\mathrm{\z}=1.59$).  Right:
Difference of in- and output magnitudes $\Delta m$ vs.\ output \V\
band magnitude of the mean relation to the left. The overplotted
errors are the 1$\sigma$ error ranges of the measured spread of host
galaxy flux, as determined from the set of simulations with different
parameter combinations and different noise realisations (see
text). The vertical lines gives the edge of the reliability region for
the \V\ band as above. For the \z\ band again the zeropoint has to be
shifted.
}
\end{figure*}

\subsection{Systematic offsets and errors}\label{sec:errors}

While the mere detection of an AGN host galaxy can be achieved with
comparably little effort, the determination of flux error bars and
systematic offsets is much more complicated. We performed extensive
simulations of artificial quasar images, composed from empirical PSFs
and host galaxy models plus artificial noise matching the actual flux
and noise distribution in real images. These simulations are described
in detail by \citet{sanc04a}. We applied the peak subtraction nucleus
removal as well as \galfit\ to a set of $\sim$2000 quasar images
created in this way. Comparing input and output parameter values
yielded mean magnitude offsets as well as statistical errors for the
individual host galaxy magnitudes (Fig.~\ref{fig:corrs_and_errors_z}).

The simulations give reliability regions and error bars. The left
panel in Figure~\ref{fig:corrs_and_errors_z} shows which magnitudes
are recovered for a given synthetic host galaxy. Since the input set
covers a large range of different morphological configurations, scale
lengths, nucleus-to-host ratios, etc., the recovered values will
scatter. Close to the detection limit, the scatter and the corrections
grow rapidly as a function of magnitude; additionally, the ability to
differentiate between different morphological types will generally be
lost. The combination of these effects is reflected in the spread of
the output of the simulations. This measured spread is a direct
estimate for the uncertainties of the total flux (right panel in
Figure~\ref{fig:corrs_and_errors_z}).

From these simulations we adopt approximate regions in brightness
where host galaxy magnitudes can be reliably determined, with
correction of 0.25 to a maximum of 0.6~mag. These regions go down to
$\mathrm{\V}=26.2$, $\mathrm{\z}=24.6$ for the peak subtraction method
and the present data. Outside the corrections and errors increase. In
three cases, marked in Table~\ref{tab:results_vz} with `?' in column
$Z_\mathrm{hg}$, the observed magnitudes extend to outside these
regions; here we continue using the derived corrections for these
three objects, but the so derived host galaxies are more uncertain and
their magnitudes should be taken with care. Notice that the \z\ band
data are substantially shallower than the \V\ band, mainly a
consequence of the ACS detector sensitivity.

\begin{figure*}
\includegraphics[clip, width=\colwidth]{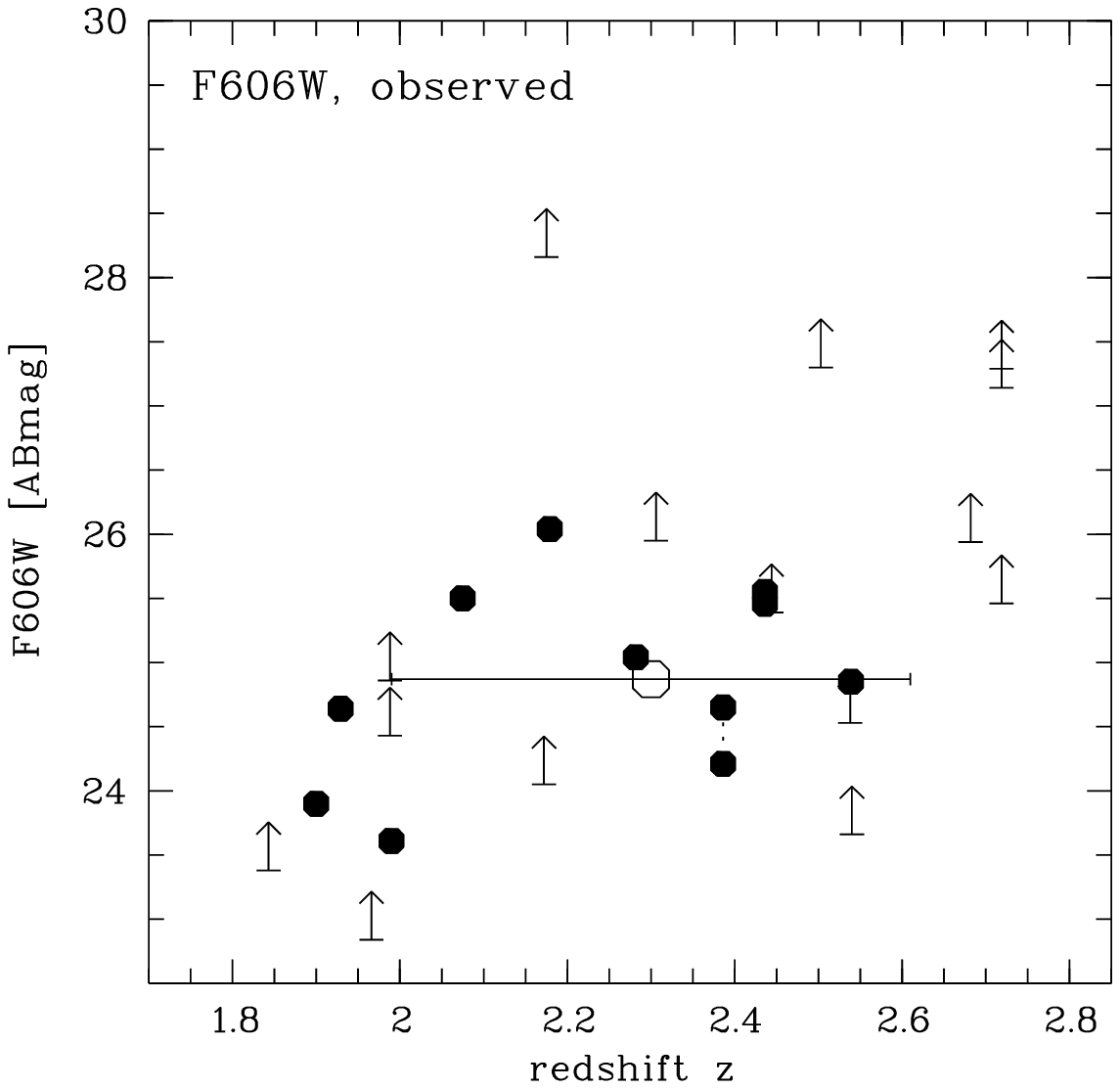}
\hfill
\includegraphics[clip, width=\colwidth]{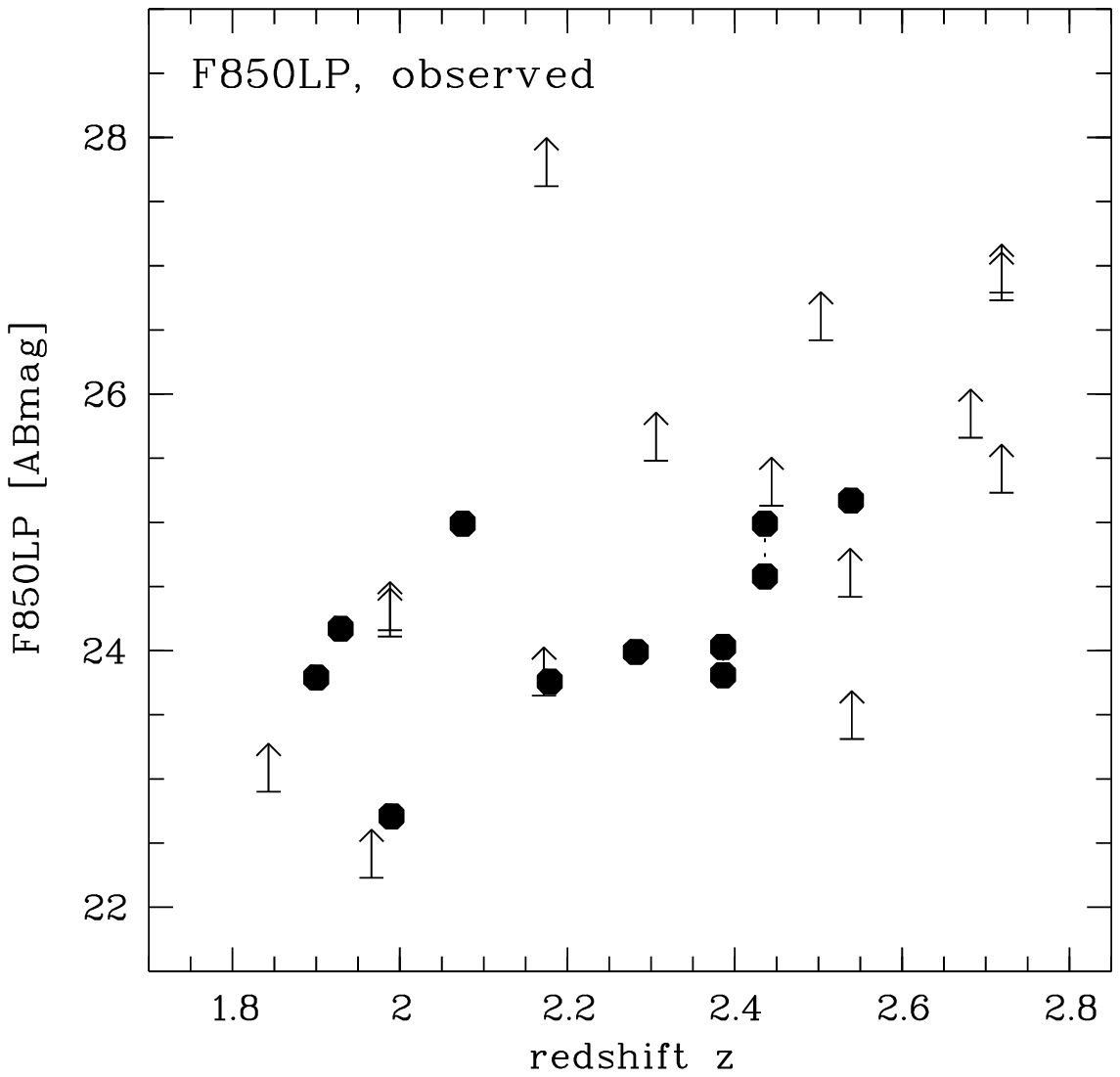}\\
\includegraphics[clip, width=\colwidth]{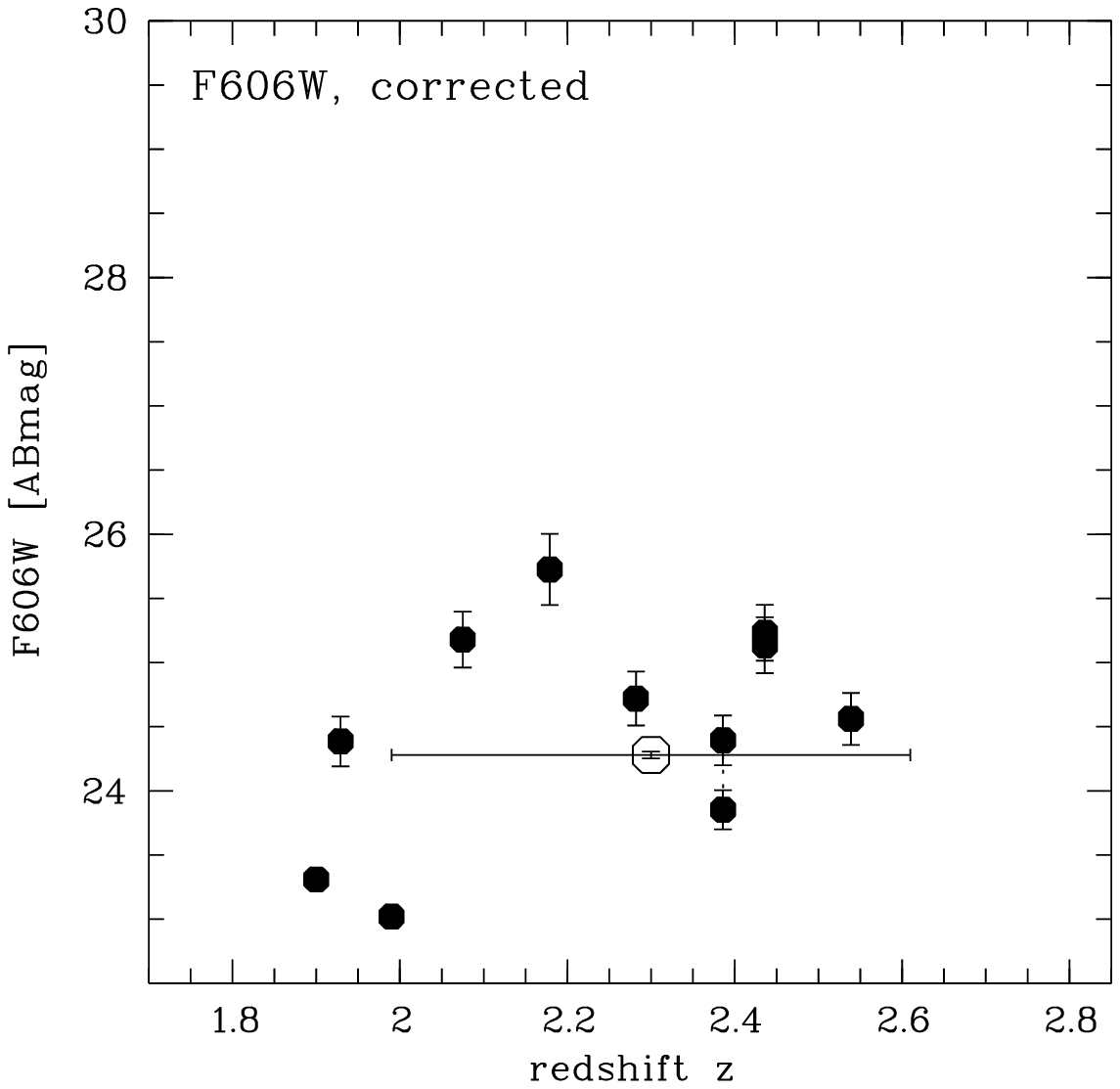}
\hfill
\includegraphics[clip, width=\colwidth]{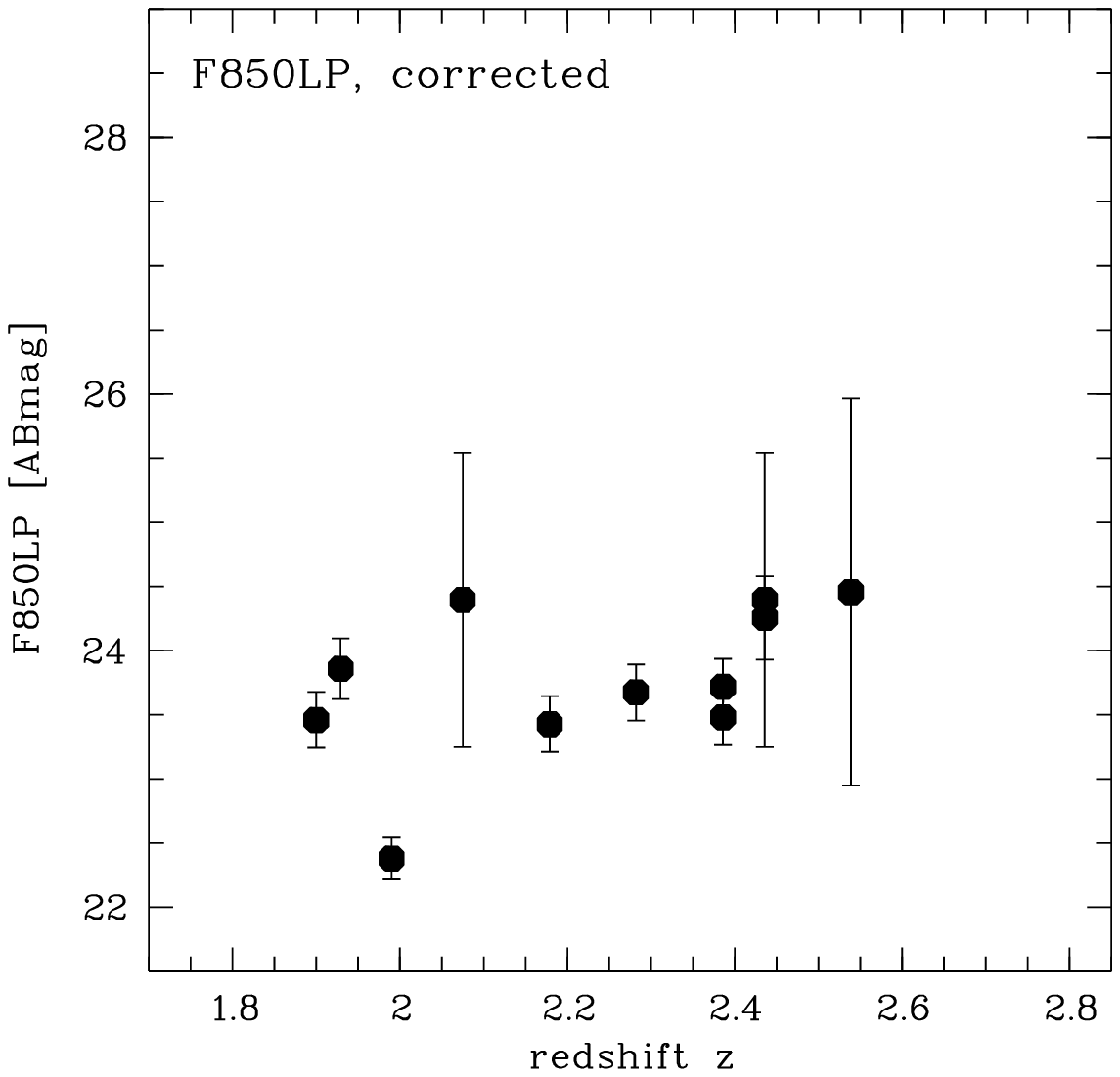}
\caption{\small \label{fig:errorsimpeak} 
Peak subtraction host galaxy magnitudes. Top: uncorrected for
oversubtraction, bottom: corrected. Left \V, right \z.  The arrows
give upper limits for the unresolved HGs, the open symbol for \V\ is
the HG extracted from the stacked unresolved AGN images. The
horizontal error bar for the stack gives the $1\sigma$ redshift range
of the coadded AGN. In the bottom diagrams the vertical errors are
uncertainties from the subtraction of the nucleus and are given as
determined from the simulations (see Sect.~\ref{sec:errors} and
Fig.~\ref{fig:corrs_and_errors_z}).
}
\end{figure*}

\section{Results}\label{sec:results}

\subsection{Detected host galaxies}\label{sec:resolved}

Using the above criteria for detecting a residual host galaxy, we find
nine of the 23 host galaxies to be resolved in both bands, although
some lie close to the sensitivity limit. One object formally fell
above the 5\% level in one band but not the other; \combo\ 33630 at
$z=2.719$ might be marginally resolved in \z\ and shows a structure at
1\arcsec distance that might be a tidal arm or a foreground
object. With this object lying at the highest redshift of the sample
we do not consider this a clear detection. As mentioned, for three
further cases in the \z\ band the host galaxies are very faint (marked
with a `?' in the $Z_\mathrm{hg}$ column in
Tab.~\ref{tab:results_vz}). While their flux is above 5\% of the
total, their raw magnitudes fell 0.8--1.0~mag outside the reliability
region where corrections and associated errors are still small. This
low S/N is also reflected in the radial profiles (see
Appendix~\ref{sec:appendix}).

As described above, tests with field stars show that 12\% of all
objects ($\sim$3 objects) might show spurious `host galaxies' at the
5\% flux level, and 3\% (0 or 1 objects) at the 10\% flux level. In
\V\ five of our objects fall with their host fluxes between these two
values. In the \z-band these are four which include the three
uncertain ones from above. According to statistics 1--3 of these might
be spurious detections. However, including or excluding these more
uncertain data points in the following analysis does not have an
influence on the conclusions drawn.

For each object the host galaxy flux was determined by simple aperture
photometry after subtraction of the scaled PSF, excluding resolved
companion objects. The radius of the aperture was matched the used
image size.

All extracted magnitudes are collected in Table~\ref{tab:results_vz},
and shown together with the upper limit for the unresolved objects in
the top panels of Figure~\ref{fig:errorsimpeak}. The extracted host
galaxy images and radial surface brightness profiles are shown in
Fig.~\ref{fig:allimages} in the appendix. To illustrate the behaviour
of true point sources, we included a selection of 24 field stars, 12
each in the \V- and \z-band (Figures~\ref{fig:stars_v} and
\ref{fig:stars_z}), that were subjected to the same PSF determination
and peak subtraction as described in Section~\ref{sec:detection}. We
plot the same profiles as for the AGN. This selection is random apart
from the fact that the 24 stars were observed on 24 different
tiles. For most of the stars there is no systematic positive residual
flux visible, as expected. The few that do show positive fluxes form
the spurious detection statistic described above.

While with deeper images or at lower redshift \citep{sanc04a} the
morphological appearance of the host galaxies can be determined, this
was generally not possible for the present data. Apart from the
extentions in \combo\ 00784, apparent residual structure visible in
the colour images shown in the Appendix is dominated by the PSF
subtraction procedure. This includes elongations and apparent
off-centering that is due to limited centering precision of the order
of 0.05--0.1 pixel (see e.g.\ \combo\ 05696 from tile 2).  Thus we
adopted the host galaxy morphological class in all cases as
`undecided', and then applied the systematic corrections for
oversubtraction provided by the simulations
(Fig.~\ref{fig:corrs_and_errors_z}).  The corrections were typically
of the order of $\sim 0.4$~mag, as documented by the columns with
`{\it cor}' subscripts in Table~\ref{tab:results_vz}. Here we also
list the estimated uncertainties resulting from our extensive
simulations. The distribution of corrected magnitudes is displayed in
the lower panels of Figure~\ref{fig:errorsimpeak}.

\begin{figure}
\includegraphics[clip,width=8.5cm]{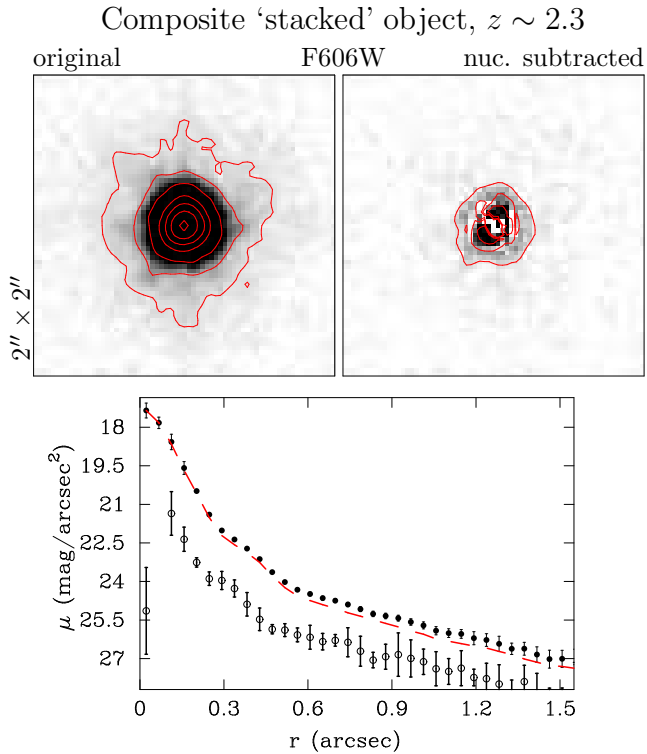}
\caption{\label{fig:unresplot}
The stacked image of individually unresolved host galaxies. Top: grey
scale original image and PSF subtracted host galaxy image (both show
$2\arcsec\times2\arcsec$) in the \V-band. On top of the linear grey
scale plot logarithmic contours with 0.5~dex spacing are
overplotted. Bottom: radial profiles of \V-band. The curves show the
upper data points with error bars from the original image, the PSF
(dashed red line) and peak subtracted host profile (lower points with
error bars from bootstrapping). The host galaxy is shown without
correction for any oversubtraction of the nucleus.
}
\end{figure}

\subsection{Detecting the undetected hosts}\label{sec:unresolved}

For 14 AGNs we find that the individual PSF-subtracted residuals are
consistent with non-detections, i.e.\ the magnitudes of individual
host galaxies lie below the 5\% limit. These objects are marked by
arrow symbols in Figures~\ref{fig:errorsimpeak} and
\ref{fig:nuchost}. It is interesting to note that these flux limits on
the hosts are by no means all outstandingly faint. For many, the
reason why they were not detected is the high contrast with the AGN
which for these objects has above average brightness. Only three
objects show non-detections that indicate exceptionally faint host
galaxies. We will further discuss the implications of the detection
limits in Section~\ref{sec:colours}.

In an attempt to assess at least the mean host galaxy properties of
the unresolved AGN, we simply coadded the images of all 14 objects,
one of which was observed in two frames. This yielded a very deep
image with effectively 15 orbits of integration time. We also combined
their PSFs, weighted by the relative flux of the AGNs, and created
combined variance and PSF rms frames. The higher S/N in combination
with the lower PSF noise (PSF and AGN position are sampled at 14
different subpixel points) yields a nominal increase in surface
brightness sensitivity of $\sim$1.5~mag. The resulting `object' has a
mean redshift of $z=2.3$ (weighted by AGN flux), and total magnitudes
of $m_{\mathrm{\V},\mathrm{tot}}=21.5$ and
$m_{\mathrm{\z},\mathrm{tot}}=21.6$, respectively.  This image is
shown in Figure~\ref{fig:unresplot}.  With the higher sensitivity we
now indeed find a host galaxy component in the \V-band image after PSF
subtraction of 4.4\% of the total flux. The radial surface brightness
profile also shows a small excess over an unresolved point source. In
both cases this flux is highly significant as we confirmed using a
bootstrap simulation for the composition of the coadded frame from the
15 frames. In the bootstrap simulation we constructed 100 new sets of
15 frames each, drawn with repetition from the original 15 frames,
coadded the images in each set and did the flux analysis as above.
The uncertainty in the total flux estimated from these 100
realisations is $\sigma=1.05$\% of the total flux, or 25\% in host
galaxy flux. All realisations yielded substantial positive
fluxes. The error is resulting from a combination of PSF uncertainty
and the noise inside the scaling aperture of 4 pixel diameter. We show
the uncertainties in the radial surface brightness determined from
bootstrapping as error bars for the derived host galaxy in
Figure~\ref{fig:unresplot}. The so extracted magnitudes for host
galaxy and nucleus in the \V-band are listed in the last row of
Table~\ref{tab:results_vz} (the `stack' object). The \z-band stack,
however, with its lower sensitivity showed a much weaker signal than
the \V-band, too faint to reliably be classified as resolved.

\begin{table*}
\begin{center}
\caption{Photometry results for the \V- (for brevity: `$V$'-) and \z-
(`$Z$') bands.}
\label{tab:results_vz}
\scriptsize
\begin{tabular}{cccccccccccc}
\tableline
\tableline
ID&Tile&
$V_\mathrm{tot}$\tablenotemark{a}&
$V_\mathrm{hg}$\tablenotemark{b}&
$V_\mathrm{hg,cor}$\tablenotemark{c}&
$V_\mathrm{nuc,cor}$\tablenotemark{d}&
N/H$_{V,\mathrm{cor}}$\tablenotemark{e}&
$Z_\mathrm{tot}$\tablenotemark{a}&
$Z_\mathrm{hg}$\tablenotemark{b,}\tablenotemark{f}&
$Z_\mathrm{hg,cor}$\tablenotemark{c}&
$Z_\mathrm{nuc,cor}$\tablenotemark{d}&
N/H$_{Z,\mathrm{cor}}$\tablenotemark{e}\\
\tableline
19965&23& 20.59& 23.9& 23.3$\pm0.05$& 20.7& 11.3& 20.29& 23.8 & 23.5 $\pm0.2$& 20.4&17.5\\
30792&82& 21.60& 24.6& 24.4$\pm0.2$ & 21.7& 12.0& 22.06& 24.2 & 23.9 $\pm0.2$& 22.3& 4.2\\
18324&19& 22.00& 23.6& 23.0$\pm0.05$& 22.5&  1.6& 21.33& 22.7 & 22.4 $\pm0.2$& 21.9& 1.6\\
05498&01& 22.91& 25.5& 25.2$\pm0.2$ & 23.1&  7.1& 22.28& 25.0?& 24.4 $\pm1.2$& 22.5& 6.0\\
51835&55& 23.01& 26.0& 25.7$\pm0.3$ & 23.1& 11.2& 22.41& 23.8 & 23.4 $\pm0.2$& 23.0& 1.6\\
00784&05& 23.28& 25.0& 24.7$\pm0.2$ & 23.6&  2.8& 22.80& 24.0 & 23.7 $\pm0.2$& 23.4& 1.2\\
05696&02& 22.73& 24.7& 24.4$\pm0.2$ & 23.0&  3.6& 22.32& 23.8 & 23.5 $\pm0.2$& 22.8& 1.9\\
05696&03& 23.14& 24.2& 23.9$\pm0.15$& 23.9&  0.9& 22.68& 24.0 & 23.7 $\pm0.2$& 23.2& 1.6\\
07671&07& 22.35& 25.6& 25.2$\pm0.2$ & 22.4& 13.2& 22.24& 25.0?& 24.4 $\pm1.2$& 22.4& 6.3\\
07671&15& 22.36& 25.5& 25.1$\pm0.2$ & 22.5& 11.9& 22.22& 24.6 & 24.3 $\pm0.3$& 22.4& 5.5\\
11922&11& 22.65& 24.9& 24.6$\pm0.2$ & 22.9&  4.8& 21.93& 25.2?& 24.5 $\pm1.5$& 22.0& 9.3\\[1ex]
stack&  & 21.47& 24.9& 24.3$\pm0.05$& 21.6& 12.3& 21.60&\multicolumn{2}{c}{unresolved} && \\
\tableline
\end{tabular}
\tablecomments{All magnitudes are uncorrected for
Galactic extinction, $A(\mathrm{\V})=0.024$ and $A(\mathrm{\z})=0.014$.}
\tablenotetext{a}{Total magnitude of the object.}
\tablenotetext{b}{Raw measured host galaxy magnitude.}
\tablenotetext{c}{Host galaxy magnitude corrected for oversubtraction.}
\tablenotetext{d}{Corrected nuclear magnitude.}
\tablenotetext{e}{Nuclear to host galaxy flux ratio.}
\tablenotetext{f}{Objects marked `?' lie outside the adopted reliability
region (see text).}
\end{center}
\end{table*}

\begin{table}
\begin{center}
\caption{Observed $(\mathrm{\V} - \mathrm{\z})$ colours and star formation rates for the resolved host
galaxies and the `stacked AGN', in case of the colours being produced
by constant star formation.}
\label{tab:colours}
\begin{tabular}{ccccc}
\tableline
\tableline
ID&Tile&$z$&$(\mathrm{\V} - \mathrm{\z})$\tablenotemark{a}&SFR \V\\
& & & &[$\mathrm{M}_\odot/\mathrm{year}$]\tablenotemark{b}\\
\tableline
19965& 23 &1.90&--0.2$\pm0.2$&  11 \\
30792& 82 &1.929&  0.5$\pm0.3$&  4 \\
18324& 19 &1.990&  0.6$\pm0.2$& 15 \\
05498& 01 &2.075&  0.8$\pm1.2$&  2 \\
51835& 55 &2.179&  2.3$\pm0.4$&  1.5 \\
00784& 05 &2.282&  1.0$\pm0.3$&  4 \\
05696& 02 &2.386&  0.9$\pm0.3$&  6 \\
05696& 03 &2.386&  0.1$\pm0.3$&  9 \\
07671& 07 &2.436&  0.8$\pm1.2$&  3 \\
07671& 15 &2.436&  0.9$\pm0.4$&  3 \\
11922& 11 &2.539&  0.1$\pm1.5$&  5 \\[1ex]
stack&    &2.3  &  $<0.0$     &  6 \\
\tableline
\end{tabular}
\end{center}
\tablenotetext{a}{Galactic extinction corrected.}
\tablenotetext{b}{Star formation rates inferred from the UV flux in
the observed \V-band.}
\end{table}

\begin{figure}
\includegraphics[clip,angle=0,width=8.5cm]{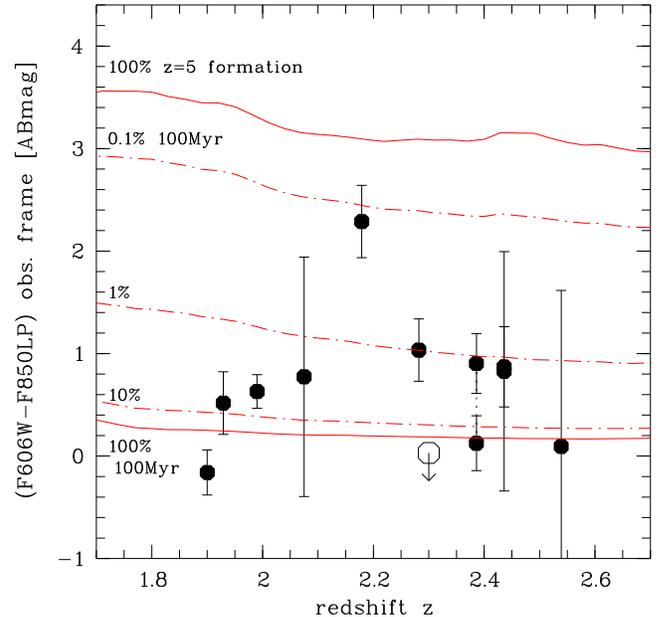}
\caption{\label{fig:SSP_colours_AB} 
Observed colours ($\mathrm{\V} - \mathrm{\z}$) of the sample from PSF
peak subtraction (circles), the open symbol marks the upper limit for
the `stacked' AGN. Overplotted are two single burst models from
\citep[solar metallicity][]{bruz03} (solid lines). The upper curve
is for a passively evolving burst at $z=5$, the lower for burst of
100~Myr age, relative to each redshift. The dot-dashed lines are
mixtures between the two, with a (from top) 0.1\%, 1\% and 10\%
fraction of mass of the 100~Myr population on top of 99.9\%, 99\% and
90\% of the $z=5$ population.
}
\end{figure}

\begin{figure}
\includegraphics[clip,angle=0,width=8.5cm]{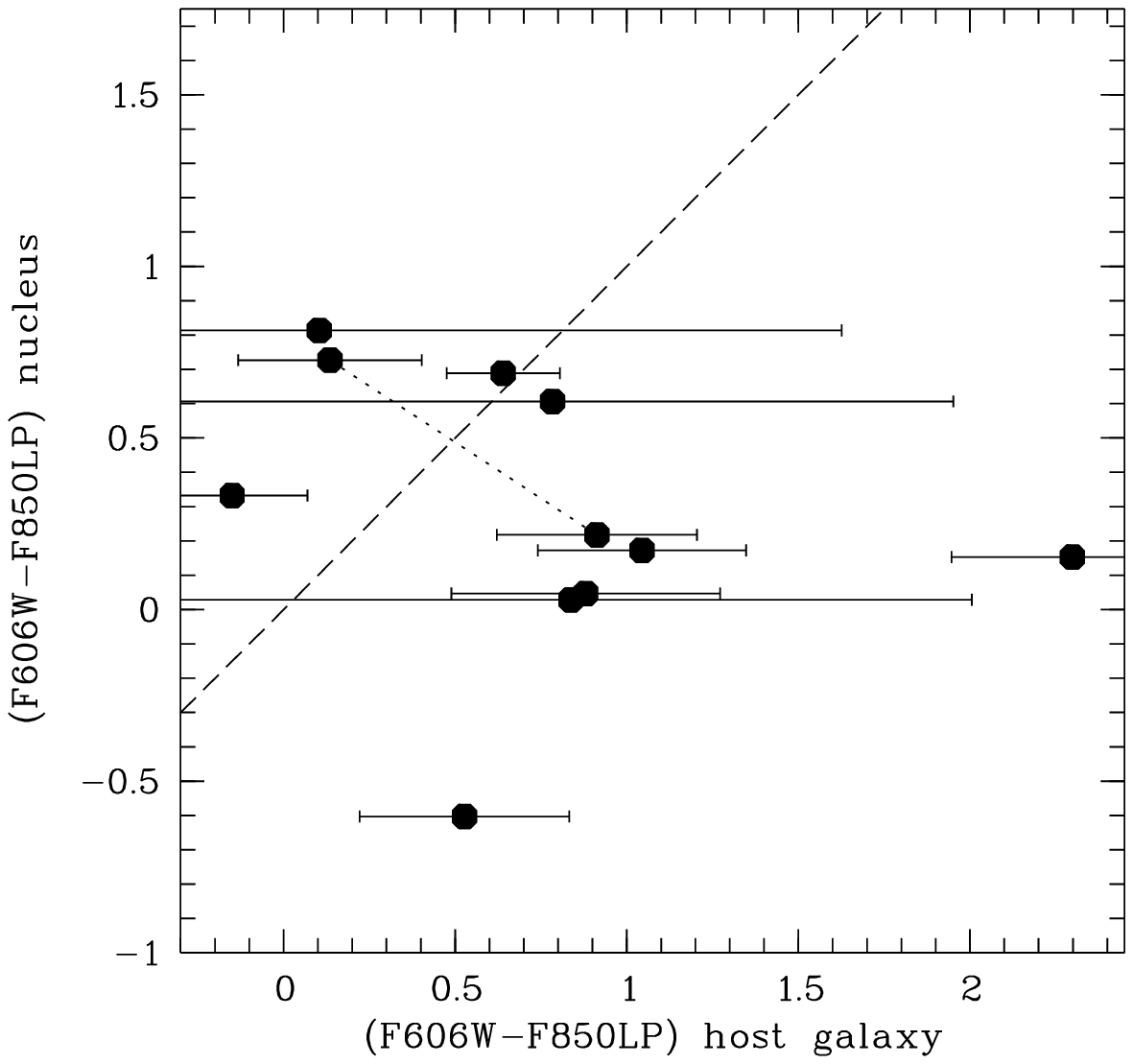}
\caption{\label{fig:nuc_host_col} 
Nuclear vs.\ host galaxy colours for the objects in the sample with
available colours. The two objects imaged twice are connected with a
dotted line (in one case not visible to near identical colours). The
dashed line is the 1:1 relation to guide the eye.
}
\end{figure}

\section{Discussion}\label{sec:discussion}

\subsection{UV colours}\label{sec:colours}

From the corrected magnitudes we have derived
$(\mathrm{\V}-\mathrm{\z})$ colours for the detected host galaxies.
These are listed in Table~\ref{tab:colours}, now including also
correction for Galactic dust extinction. However, with $E(B-V)=0.008$
\citep{schl98} these values of $A(\mathrm{\V})=0.024$ and
$A(\mathrm{\z})=0.014$ negligibly affect the colours.

At these redshifts, the observed photometric bands correspond to the
rest frame ultraviolet, ranging from 2160\,\AA\ at $z=1.8$ to
1616\,\AA\ at $z=2.75$ in \V\ and 3150\,\AA\ to 2350\,\AA\ in \z,
respectively, so for this redshift range pure rest-frame UV colours
are observed.  Figure~\ref{fig:SSP_colours_AB} shows the measured
values plotted against $z$.  There is no discernible colour trend with
redshift. All points fall within a relatively narrow range of colours;
apart from \combo~51835, the objects occupy a band of
$-0.2<(\mathrm{\V}-\mathrm{\z})_\mathrm{observed}<1.0$.  The open
symbol represents our stacked `average' AGN constructed from the 14
unresolved objects. Although it was not resolved in the \z\ band, the
upper limit on its colour is actually consistent with the values
derived for several of the detected objects.

One critical issue in measuring UV luminosities of barely resolved AGN
hosts is the lingering possibility that flux may have spilled over
from the nuclei. This could have happened as a purely observational
artefact due to imperfect PSF removal, or physically by scattering of
UV photons off dust in the the host galaxy. The latter phenomenon is
known to be relevant in high redshift radio galaxies
\citep{vern01}. Independently of the underlying mechanism, any such
cross-contamination should be visible in a correlation of host galaxy
with nuclear colours. Figure~\ref{fig:nuc_host_col} shows these
colours plotted against each other. No correlation is visible and we
conclude that a substantial contamination of the host galaxy light
from the AGN is very unlikely. Notice also that when considering
physical scattering in the hosts, powerful radio galaxies are huge
massive entities of generally vastly different appearance compared to
the relatively modest AGN hosts featuring in our sample.

\begin{figure}
\includegraphics[clip,angle=0,width=8.5cm]{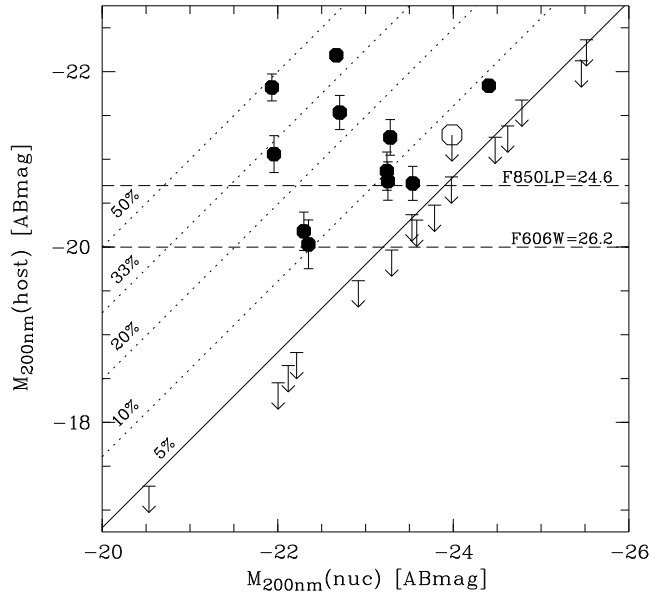}
\caption{\label{fig:nuchost} 
Nuclear vs.\ host galaxy absolute magnitude at 200~nm. Shows are
objects with resolved host galaxies (solid symbols), the coadded
stacked object (open circle with upper limit arrow) and upper limits
for objects with unresolved host (arrows). The diagonal lines mark
positions of constant fraction of host galaxy light of 5\%, 10\%,
20\%, 33\% and 50\% of the total light, assuming a constant spectral
slope $\beta$. The horizontal dashed lines show the magnitude of host
galaxies at the $\mathrm{\V}=26.2$~mag and $\mathrm{\z}=24.6$~mag
detection limit, at mean redshift and mean $\beta$.
}
\end{figure}

UV colours can be converted to UV spectral slope $\beta$ independent
of redshift, when assuming that the SED can be described in the form
$F_\lambda \propto \lambda^\beta$. With $\beta$ known, the absolute
magnitude at 200~nm, $M_\mathrm{200nm}$, can be computed directly from
the \V- or \z-band apparent magnitudes. We do this for both the host
and the nucleus and these values are shown for the full sample
including upper limits in Figure~\ref{fig:nuchost}, which illustrates
the two main constraints for resolving a host galaxy, apart from
compactness (see Sect.~\ref{sec:unresolved}). The reliability of host
galaxy photometry is constrained by S/N, thus dependent on the filter
which have different depths. This is marked by the horizontal dashed
lines which show the $M_\mathrm{200nm}$ magnitude of host galaxies at
the $\mathrm{\V}=26.2$~mag and $\mathrm{\z}=24.6$~mag edges of the
adopted regions of reliability (see Section~\ref{sec:errors}), at mean
redshift and mean $\beta$.

As the second effect the maximum nucleus--host contrast appears as the
scatter of the unresolved objects around a line shifted by 3.2~mag
from unity, corresponding to 5\% of the total flux. Here the scatter
is only induced by the assumption that all unresolved host galaxies
have a the $\beta$ value the upper limit determined for the stacked
image. The diagonal dotted lines mark lines of 10\%, 20\%, 33\% and
50\% of the total flux associated with the host galaxy. Thus in total
the region right of the solid diagonal line is inaccessible to host
galaxy detection with the current data and method of analysis. We
would like to emphasise that the similarity of nuclear properties for
resolved and unresolved host galaxies suggests that also the host
galaxy properties are similar -- thus supporting that the data point
for the coadded stacked object is not far off the individually
resolved objects in all plots.

\begin{figure}
\includegraphics[clip,angle=0,width=8.5cm]{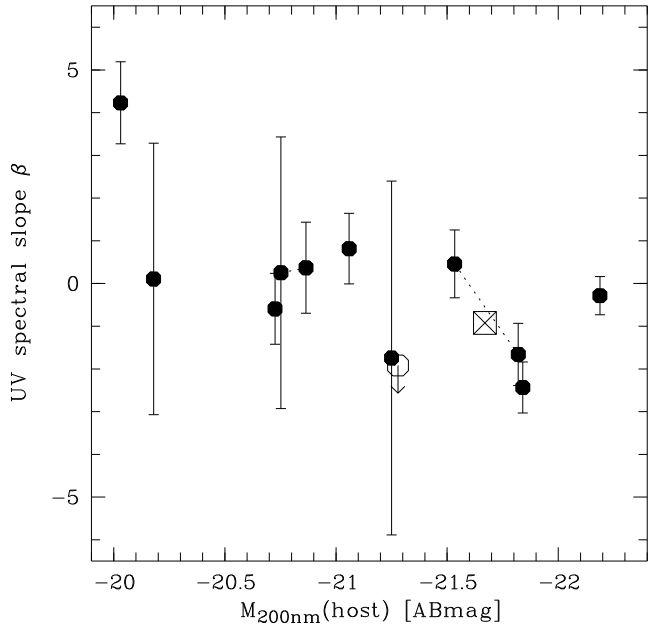}
\caption{\label{fig:beta} 
Spectral slope $\beta$ in the UV ($F_\lambda \propto \lambda^\beta$)
vs.\ 200~nm absolute magnitude, uncorrected for dust. The open symbol
marks the `stacked' object, the crossed square is the mean value
obtained by \citet{shap03} for 794 Lyman break galaxies at $z\sim3$.
}
\end{figure}

In Figure~\ref{fig:beta} spectral slope is plotted against
$M_\mathrm{200nm}$ for the host galaxies, compared to the mean value
for a sample of 794 Lyman break galaxies at redshift $z\sim3$
\citep{shap03}, all values uncorrected for the influence of dust in
the galaxies.

The slight anticorrelation that seems to be visible between $\beta$
and UV luminosity suggest that the more luminous host galaxies are
having the steeper spectral slopes or bluer colours. This would mean
that more luminous host galaxies had bluer colours and thus more UV
light from young stars. To test this we computed the Spearman
rank-order coefficient for this data set. The test gave a probability
for the zero hypothesis -- uncorrelated data -- of 9\%. Thus the zero
hypothesis can not even be rejected on a 2$\sigma$ level and thus the
anticorrelation is not significant. 

A few of the most UV-luminous host galaxies fall into the same
magnitude--colour regime as the LBGs, while a number of objects are
substantially redder. Thus we do not find a positive correlation of
the amount of host UV light and luminosity.

\subsection{Origin of the host galaxy UV flux}
The top solid line in Figure~\ref{fig:SSP_colours_AB} shows the
theoretical colour of a galaxy that formed all its stars at $z=5$ and
evolved passively afterwards. This colour was computed from single
stellar population (SSP) models taken from \citet{bruz03}. The
assignment of solar metallicity in this context is arbitrary and
subject to discussion (see below). However, qualitatively there is no
strong dependence on metallicity, the exact formation redshift or the
particular choice of IMF for a given model family (\citealt{salp55} or
\citealt{chab03}) in this wavelength range.

This comparison shows that the measured UV colours of our detected AGN
hosts as well as of the stacked `mean' host galaxy are markedly bluer
than expected for an `old' population at that epoch, i.e.\
$t_{age}\sim t_\mathrm{Hubble}$. Any correction for dust in the host
galaxies will strengthen this result. Thus the blue light detected in
these galaxies must come from relatively young stars. These stars
could be forming continuously, in which case the UV luminosities can
be interpreted as indicators of the star formation rate in these
galaxies. We follow up on this option in the next subsection.
Alternatively, the UV flux could be the afterglow of a past
starburst. This is the option we consider first.

\subsubsection{Recent starburst}\label{sec:starburst}

Given the unknown dust absorption and metallicities in the host
galaxies, a single UV colour is insufficient for performing a detailed
age dating of starburst or a decomposition of stellar populations.
However, we want to compare to the available theoretical colour range
spanned by galaxy formation at $z=5$ and a very recent (100~Myr)
starburst to illustrate how mixing of an old underlying population
with most of the mass and a recent starburst influences the colour. In
Figure~\ref{fig:SSP_colours_AB} the solid lines mark these
extremes. In between these the dot-dashed lines show how contribution
of a 100~Myr component would influence the colour of an otherwise old
population. From top to bottom 0.1\%, 1\% and 10\% in mass are added
to the old population. There is a strong degeneracy between the choice
of burst age and the mixing ratio. For 10~Myr less than one tenth in
mass is required to produce the same UV colour compared to 100~Myr. If
we choose 100~Myr as a timescale similar to the dynamical timescale in
galaxies, the masses involved in that starburst would be of the order
of a few percent of the total stellar mass.

For the assumption of only one single-aged population we can rule out
a very high formation redshift -- $z=5$ corresponds to ages of
3.5--2.5~Gyr at $z=1.8$--$2.5$ in the chosen cosmology. For the
adopted set of models, the resulting age estimates would range mostly
between $\sim$0.1 and $\sim$0.7~Gyr. 

We note that the colour tracks in Figure~\ref{fig:SSP_colours_AB} are
largely flat over the redshift range of interest, and that the two
pure population and the different mixing ratios correspond to
different colours almost independently of $z$. Since our measured
colours are all quite similar, we conclude that the luminosity
weighted ages of the the UV-dominating stellar population must be
rather similar, unless younger ages and more reddening in some objects
conspire.

Clearly, a single UV colour is insufficient to perform a reliable age
dating, with all broad band colours being affected by various
degeneracies with respect to dust and metallicities. However, we have
reason to believe that at least the central lines of sight towards the
AGN are reasonably free of dust extinction (because the AGN sample is
selected by optical/UV flux), and we therefore do not expect dust to
play a major role. At any rate, significant dust extinction would make
the host galaxies intrinsically bluer than what we observe.  On the
other hand, assuming a metallicity lower than solar would shift all
curves in Figure~\ref{fig:SSP_colours_AB} downward, resulting in older
age estimates. Reducing the metallicity to $Z = 0.004$ (1/5 solar)
gives an single burst age increase by a factor of two (for Bruzual \&
Charlot models). We conclude that if the UV light in our host galaxies
is emitted by a passively evolving population of young stars, this
population is typically much younger than a Gyr.

The diagram in Figure~\ref{fig:nuchost} shows that no correlation
exists between the amount of stellar UV light from the host galaxies
and the amounts of UV radiation produced by the nuclei. If the latter
is taken as a measure of the amount of matter accreted by the nuclei
then in the context of a recent starburst the size of the starburst
and the amount of accreted matter must be governed by different
mechanisms. If the accretion rate is primarily defined by the nuclear
mass, and a correlation between galaxy and black hole mass is assumed,
then the size of the starburst is independent of the bulk stellar mass
of the host galaxy. Other factors must be dominating the amount of
(gas) mass involved in the starburst. This can be either the total
amount of gas available in the galaxy, the size of the region involved
in the starburst, the strength of an interaction responsible for the
starburst, etc. In any case the amounts of gas involved are variable
for a given nuclear luminosity. Including the upper limits in
Figure~\ref{fig:nuchost} the host luminosities span e.g.\ $\sim$4~mag
at $M_{200~\mathrm{nm}}(\mathrm{nuc})= -22$, i.e.\ the amount of gas
involved can vary by a factor of $\sim40$ or more.

\begin{figure}
\includegraphics[clip,angle=-90,width=8.5cm]{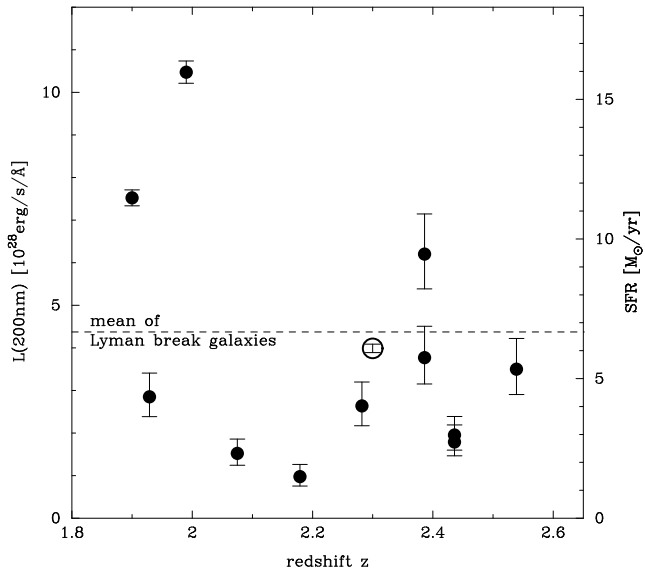}
\caption{\label{fig:sfr} 
Rest frame 200nm luminosities, and star formation rates as derived
from from the \V-band, both uncorrected for dust. The open symbol
marks the SFR of the `stacked' object created from the AGNs with
individually unresolved host galaxies. The horizontal dashed line is
the value obtained by \citet{erb03} for Lyman break galaxies at
$z=2.5$.
}
\end{figure}

\subsubsection{Estimating a host star formation rate}\label{sec:sfr}

We now interpret the detected UV emission in the alternative framework
of being due to young stars forming continuously in the AGN host
galaxies. Under this assumption it is possible to estimate the star
formation rate (SFR) of the host galaxies from the measured rest frame
UV luminosities. Following \citet{kenn98} and using the conversion of
AB magnitudes into monochromatic fluxes we obtain
\[
\mathrm{SFR}\left(\frac{\mathrm{M}_\odot}{\mathrm{year}}\right) = 1.8 \times10^{-27} 
\left(\frac{d_l^2 10^{-0.4\left(m_\mathrm{AB} + 48.6\right)}}{1+z}\right)
\]
where $d_l$ is the luminosity distance to the AGN in cm, and
$m_\mathrm{AB}$ is the observed UV magnitude at an arbitrary
wavelength between 1500--2800\,\AA .  With this formula we can now
convert our \V\ band luminosities (which are much deeper than the \z\
band data) into star formation rates. As long as intrinsic dust
attenuation is neglected, these values are of course mere lower
limits.

The resulting SFR values are listed in Table~\ref{tab:colours} and
generally amount to a few solar masses per year, with remarkably
little variation. The maximum value found is
$\sim$15~$\mathrm{M}_\odot \:\mathrm{yr}^{-1}$, and the mean is
$\sim$6~$\mathrm{M}_\odot \:\mathrm{yr}^{-1}$, including the results from
the stacked image of individually unresolved host galaxies. This value
is thus representative for the full sample of 23 AGN.  We show the
distribution of star formation rates vs.\ redshifts in
Figure~\ref{fig:sfr}. No trend emerges, consistent with the previously
established observations that neither colours nor luminosities show
any significant trend with $z$. 

This is again compared to the star formation rates for LBGs at
$z=2\ldots2.6$ from \citet{erb03}, as determined from the UV flux and
uncorrected for dust. There are a few host galaxies with higher UV
flux while for the majority it is smaller by a factor of 2--3.

However, there is a principal caveat in our guiding assumption of this
subsection: The UV light from galaxies with strong ongoing star
formation is expected to be totally dominated by the youngest
stars. \citet{kenn98} pointed out that the spectral shape of galaxies
with a constant SFR over at least $\sim 100$~Myr is basically flat (in
$f_\nu$) between 1500--2800\,\AA , assuming a \citet{salp55} IMF.
This would lead to an expected UV colour for our objects of
$(m_\mathrm{\V} - m_\mathrm{\z})=0$ or spectral index $\beta=-2$, more
or less independently of redshift (within the $z$ range of our
sample), inconsistent with our observations for most objects (see
previous subsection, Figure~\ref{fig:SSP_colours_AB} and
Table~\ref{tab:colours}). In other words, most detected host galaxies
of our sample have UV colours that are too `red' for a simple
continuous star formation scenario, while the stacked host galaxy is
roughly consistent. If for the resolved hosts the \z\ band fluxes were
used instead of the \V\ band, SFR values would be higher by roughly a
factor of 2.

This apparent inconsistency could be resolved in several ways. The
initial assumption could be wrong, and the UV flux originates not from
freshly formed stars, but from a passively evolved starburst as
outlined in Sect.~\ref{sec:starburst}. The star formation rate might
not have been constant over the past, so that varying amounts of stars
of different masses and ages would have been formed; such a
configuration is always possible, and our only argument against it
would invoke Occam's razor. Finally, as discussed in the previous
section, there could be an underlying older stellar population
contributing more to \z\ and less to \V\ (but note that by sample
design, also \z\ is completely below the Balmer jump at all relevant
redshifts). This is the scenario that is favoured by our intermediate
redshift data from \gems\ \citep{sanc04a}.

Clearly, in this study our current data set of just two UV bands is
insufficient to settle this ambiguity. However, in all three cases
there would be no continuous star formation as inferred for LBGs. The
blue light would be a result not of a continuous process but of one or
more events in the past of the host galaxy that triggered star
formation. Whether galaxy interaction or merger incidents were
responsible or the formation of bars or spiral arms is involved can
not be investigated with the present set of data.

\subsection{Comparison with other AGN host galaxy studies}\label{sec:other}

Even at low redshifts, colour data of AGN host galaxies are relatively
scant, except for rather low-luminosity AGN where the host galaxy can
be separated with relative ease.  In those cases, colours were
generally found to be consistent with morphological types, in
particular for the prevailing disk-type host galaxies
\citep{koti94,scha00}. However, when higher nuclear luminosities are
observed (which generally correlate with a larger bulge component),
then a tendency towards abnormally blue colours and younger stellar
populations emerges \citep{kauf03,jahn04a}. This tendency appears to
hold also at intermediate redshifts, as demonstrated in our companion
\HST\ paper \citep{sanc04a}, where we find more than half of the
investigated host galaxies to have bluer colours than what would be
expected from their morphological types. However, we can not confirm
increasing amounts of blue light from young stars with increasing host
luminosity as found by \citep{kauf03}. Our data are consistent with
constant UV flux for all hosts.

At high redshifts ($z \ga 2$), colour information is available only
for a handful of objects. In a ground-based study of six bright
radio-loud quasars, \citet{lehn92} found indications that the hosts
were very blue, actively star-forming galaxies. For three high
luminosity quasars, one radio-loud and two radio-quiet, \citet{aret98}
claimed very luminous envelopes and star formation rates of several
hundreds solar masses per year. More recently, \citet{hutc02}
presented optical \HST\ observations of three radio-loud and two
radio-quiet quasars and found less extreme, but qualitatively similar
results.

Because of their high radio power and optical luminosities, many
of the quasars observed in previous studies 
are probably quite different from the moderate-luminosity hosts of 
moderate-luminosity radio-quiet AGN that we focus on.
It is nevertheless interesting to see that the presence of
a significantly enhanced UV continuum and young stars seems to 
be at least qualitatively similar between QSOs of intermediate and 
high luminosities.

\section{Conclusions}   \label{sec:conclusions}

We performed the hitherto largest study of host galaxies properties of
a complete sample of high-redshift AGN. We detected the hosts and
extract colour information in 9 of the 23 AGN, and we also achieved a
statistical detection of the host in the remaining 14 from a stack
analysis. The UV luminosities can be interpreted in three ways: either
as contribution from a passively evolving population of relatively
young stars, forming typically 0.5~Gyrs ago, as a mix between a
population of old (e.g.\ $t_{age}\sim t_\mathrm{Hubble}$) stars and a
small contribution of a recently formed young population (e.g.\
0.1\%--10\% in mass at an age of 100~Myrs or 1/10th of this for age
10~Myrs), or as an indicator of ongoing star formation at a level of
$\sim$2--15\,$\mathrm{M}_\odot\:\mathrm{yr}^{-1}$ (uncorrected for
internal dust attenuation). While the first possibility is very
simplistic and appears unphysical, the UV colours actually favour the
two burst interpretations; but the possibility of on-going star
formation cannot be completely ruled out from our data.

In the framework of combined old and young populations, it is
remarkable how similar the host galaxy colours are within the sample,
and, unless different mass--age combinations conspire, hence the
estimated stellar mass fractions and ages. The derived young
population mass fractions and ages are also very similar to the values
estimated in our companion \gems\ study of AGN at $z\la 1$
\citep{sanc04a}, where we find abnormally blue rest-frame $U-V$
colours for a substantial fraction of host galaxies, particularly the
most luminous AGN in the sample. Even more, these colours and ages are
in turn very close to the mean values obtained from our ground-based
low-$z$ multicolour sample \citep{jahn04a}. While the stellar
population diagnostics of \citet{kauf03} are not immediately
convertible into our simple colour indicators, their impressive and
highly significant results point in exactly the same direction.

While the results from all these redshift regimes are similar and
point to a connection of nuclear activity and the presence of young
stars, and that mass fractions of young stars are similar, we always
find that a larger range of absolute masses is involved, showing as a
range in UV luminosity. Here we find a variation of a factor of
$\ga$40 for a given nuclear luminosity.

Our host galaxy colours span a range that reaches the colours of Lyman
Break Galaxies for a few very luminous hosts, while, as mentioned, the
colours of the majority are somewhat redder than these. A comparison
of optical/UV properties to the general population of high redshift
galaxies would be very illuminating, but large statistical samples
only become available in the near future, e.g.\ from the \goods\
project.

This persistent trend to find AGN to be associated with blue stellar
colours is intriguing and suggests a close connection between enhanced
star formation and nuclear activity. Additional support for such a
connection comes from the detection of submm CO emission in a number
of extremely luminous high-redshift QSOs and radio galaxies
\citep{omon03}, although the current sensitivity of submm telescopes
is insufficient to perform this test for less luminous AGN at high
$z$.

While the fact that there is a relation can hardly be denied, its
physical origin remains obscure. Is the enhancement of star formation
a prerequisite for nuclear activity? Is it a simultaneously occurring
phenomenon, caused by the same trigger? Or is it a consequence of the
AGN? Galaxy merging and interaction are clearly two possible
candidates to connect these two phenomena, but neither the only ones
nor are the involved physics understood. Much additional data will be
required, in particular those helping reliably to reconstruct the star
formation history in high-redshift galaxies, before any firm
conclusions can be drawn.

\acknowledgements 

Based on observations taken with the NASA/ESA {\it Hubble Space
Telescope}, which is operated by the Association of Universities for
Research in Astronomy, Inc.\ (AURA) under NASA contract NAS5-26555.
Support for the GEMS project was provided by NASA through grant number
GO-9500 from the Space Telescope Science Institute, which is operated
by the Association of Universities for Research in Astronomy, Inc.,
for NASA under contract NAS5-26555.  EFB and SFS ackowledge financial
support provided through the European Community's Human Potential
Program under contract HPRN-CT-2002-00316, SISCO (EFB) and
HPRN-CT-2002-00305, Euro3D RTN (SFS).  CW was supported by a PPARC
Advanced Fellowship. SJ acknowledges support from the National
Aeronautics and Space Administration (NASA) under LTSA Grant
NAG5-13063 issued through the Office of Space Science.  DHM
acknowledges support from the National Aeronautics and Space
Administration (NASA) under LTSA Grant NAG5-13102 issued through the
Office of Space Science.

\bibliographystyle{apj}

\appendix
\section{AGN images and surface brightness plots of AGN and stars}\label{sec:appendix}
Figure~\ref{fig:allimages} shows plots for each of the nine resolved
object plus the composite `stacked' object. Two objects appear twice
as they appear in overlapping areas of \gems\
tiles. Figures~\ref{fig:stars_v} and \ref{fig:stars_z} show a random
selection of isolated stars used to show the zero case of a point
sources without any host galaxy contribution, for comparison purposes.

\begin{figure*}
%
%
\epsscale{1.0}
\plotone{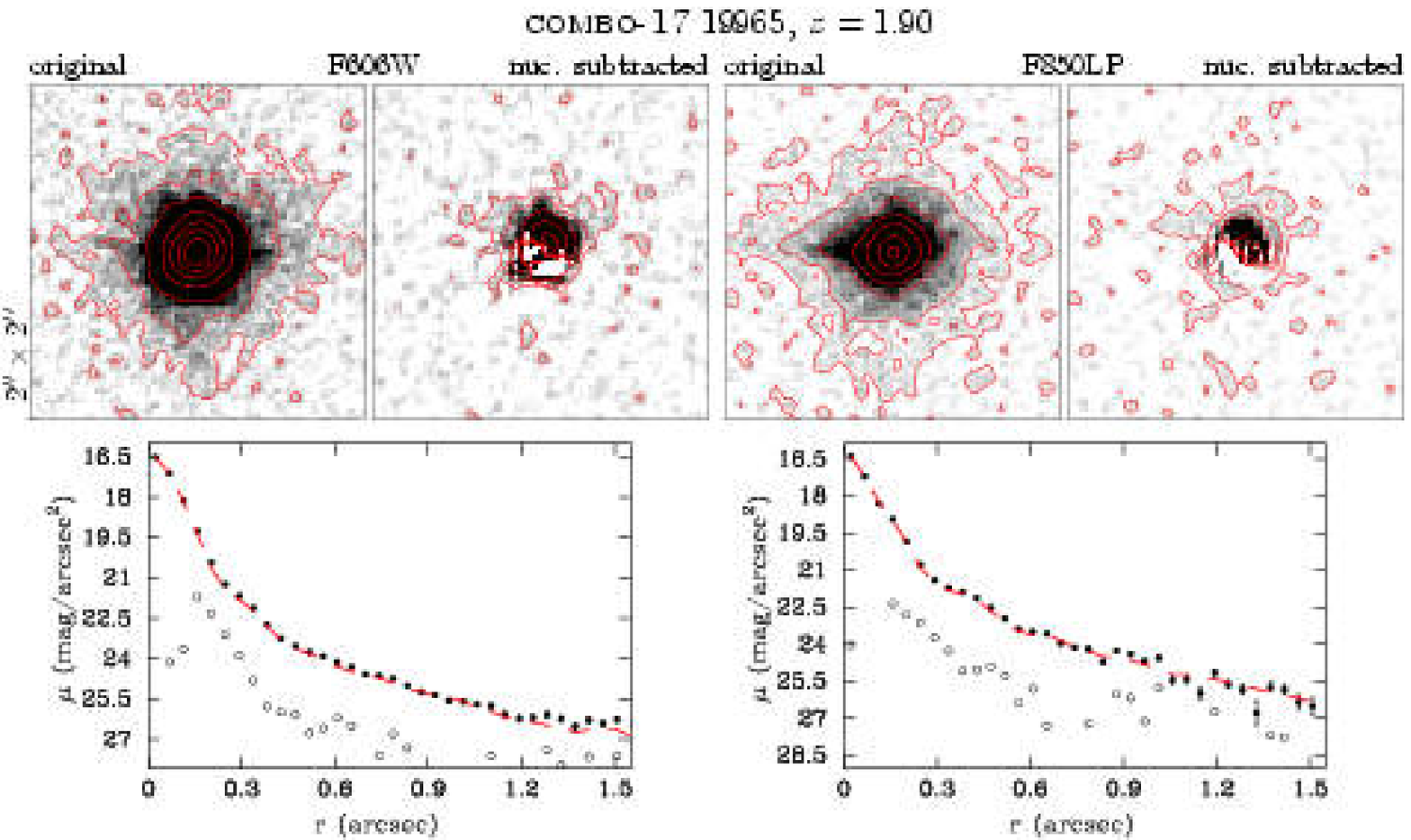}
\plotone{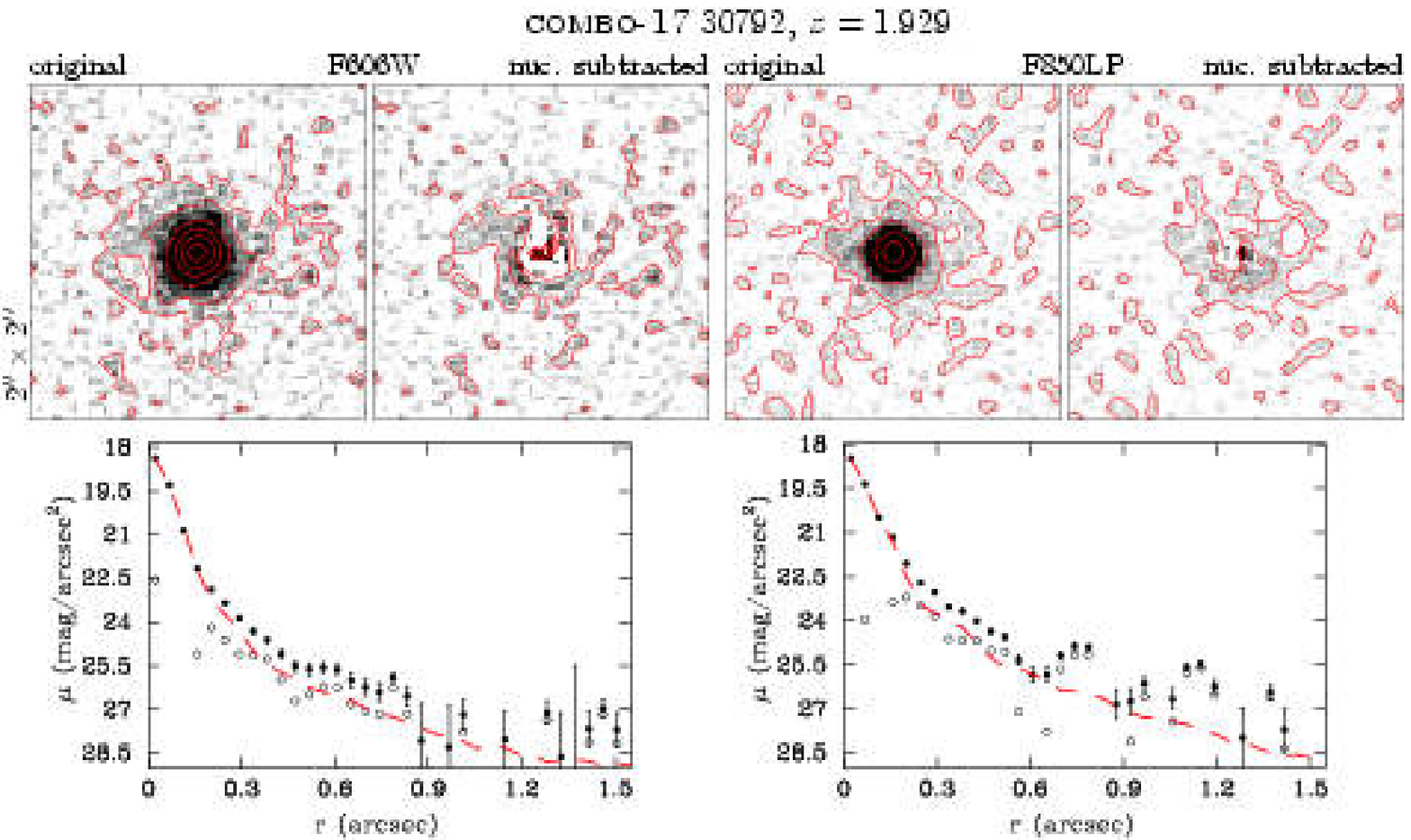}
\caption{\label{fig:allimages}
For each object the original images is shown and the uncorrected peak
scaled PSF subtracted images, for both the \V\ and \z\ filter. The
image size is 2\arcsec$\times$2\arcsec and the images are shown in a
linear grey scale, with overplotted logarithmic isophotes of 0.5~dex
spacing.  The plots below show the radial surface brightness in the
two filter, respectively. The curves show the data points with error
bars from the original image (filled symbols), the PSF (dashed red
line) and peak subtracted host profile (open symbols). The host galaxy
is shown without any correction for oversubtraction as applied to the
derived magnitudes.
}
\end{figure*}

\begin{figure*}
%
%
\epsscale{1.0}
\plotone{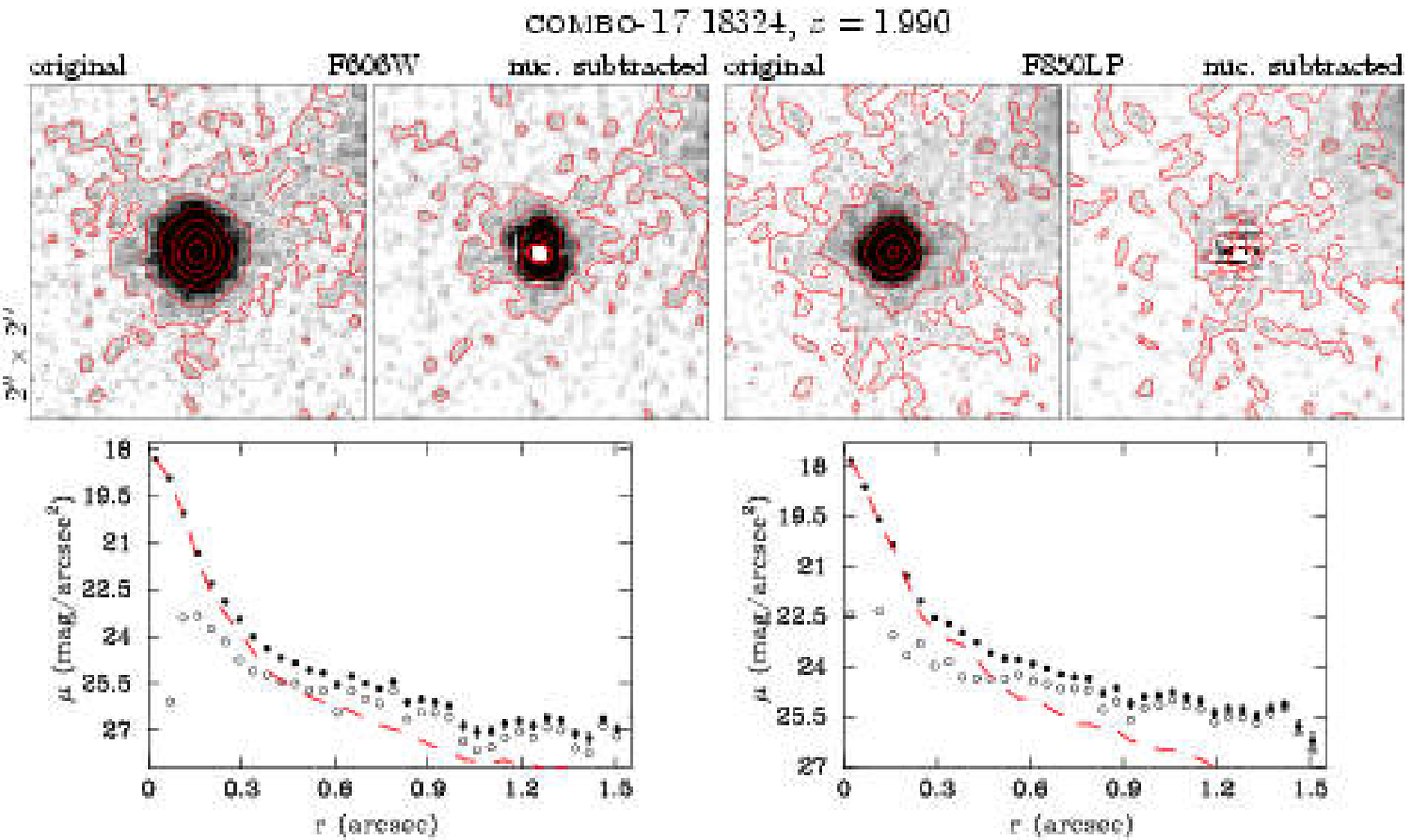}
\plotone{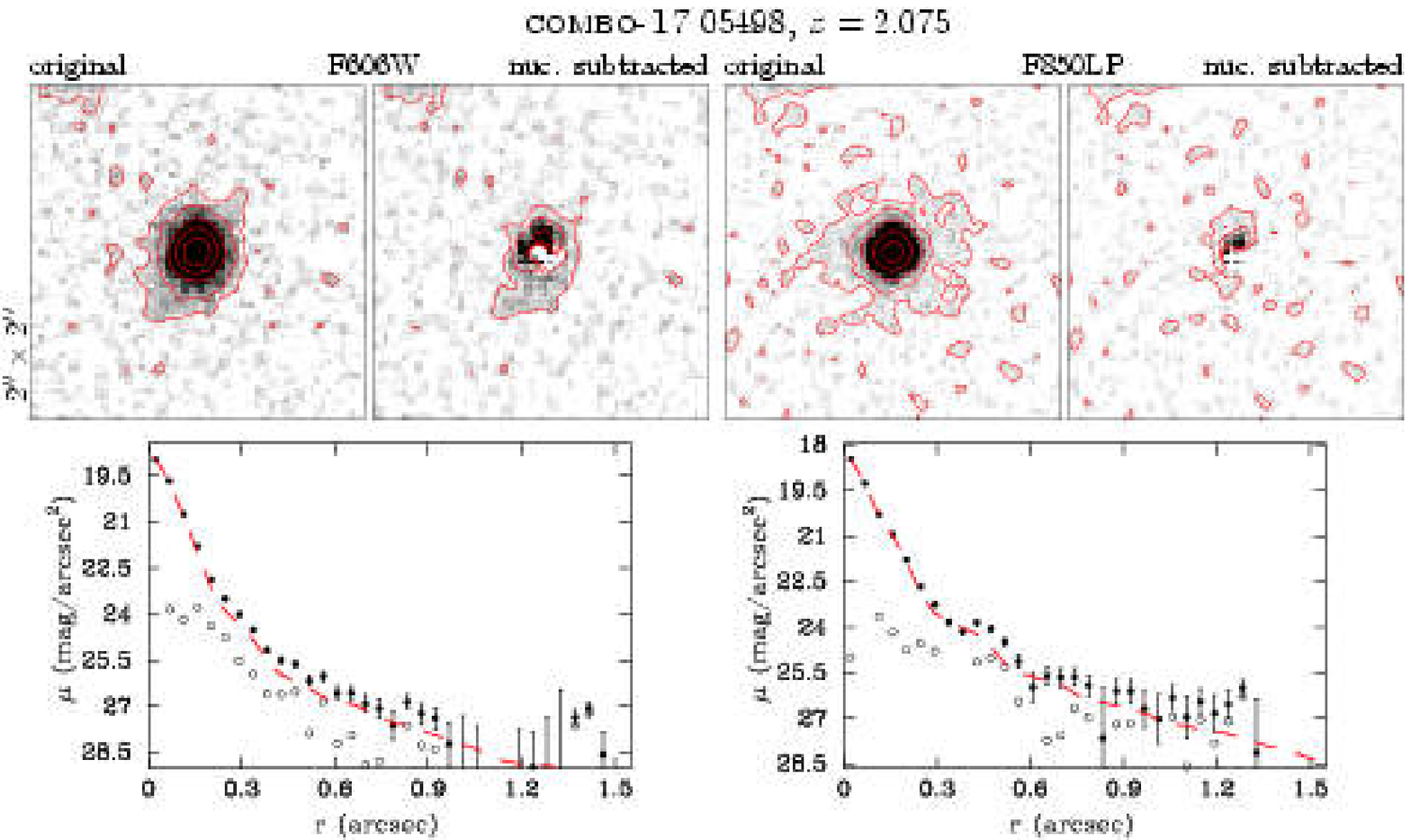}
\addtocounter{figure}{-1}
\caption{
continued
}
\end{figure*}

\begin{figure*}
%
%
\epsscale{1.0}
\plotone{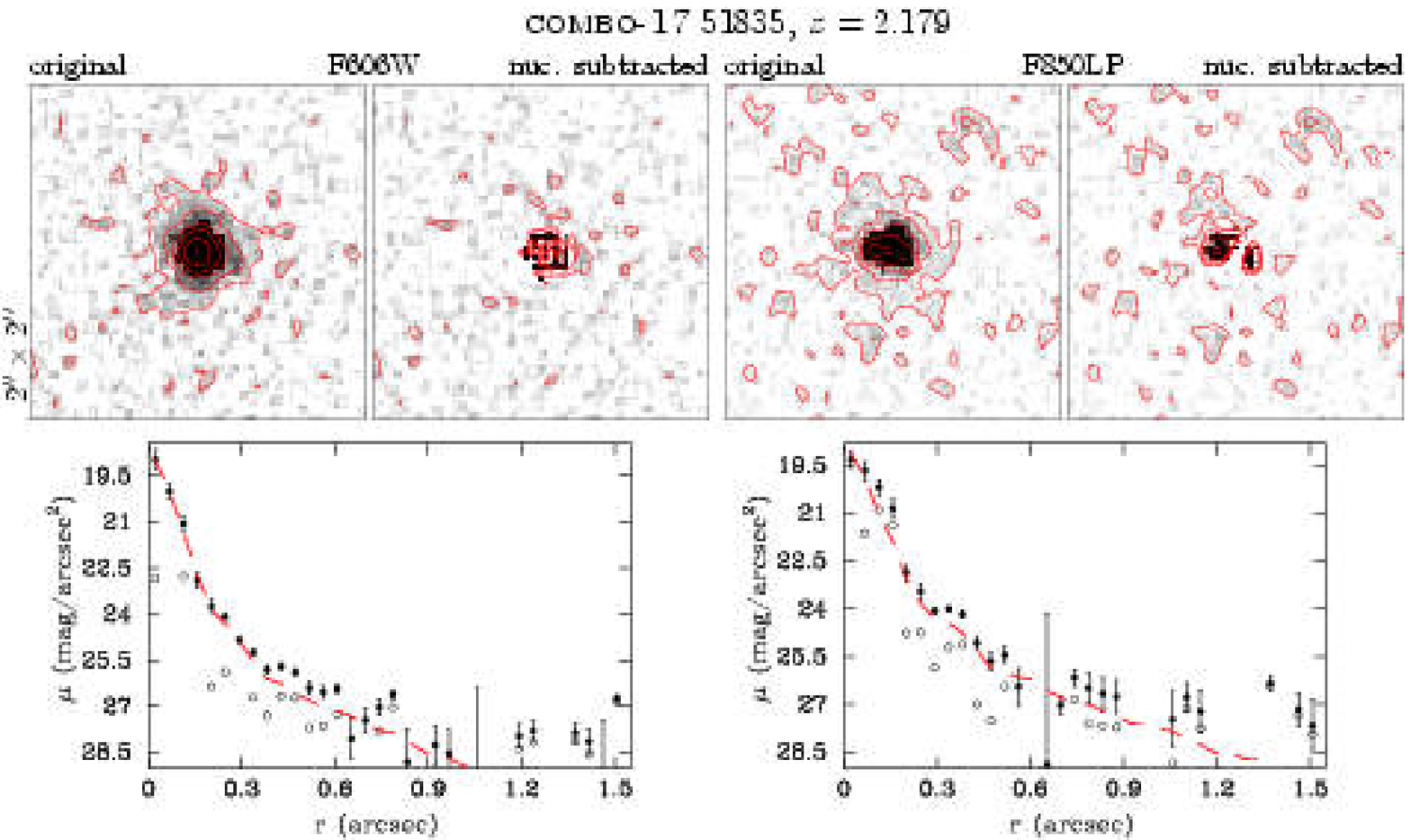}
\plotone{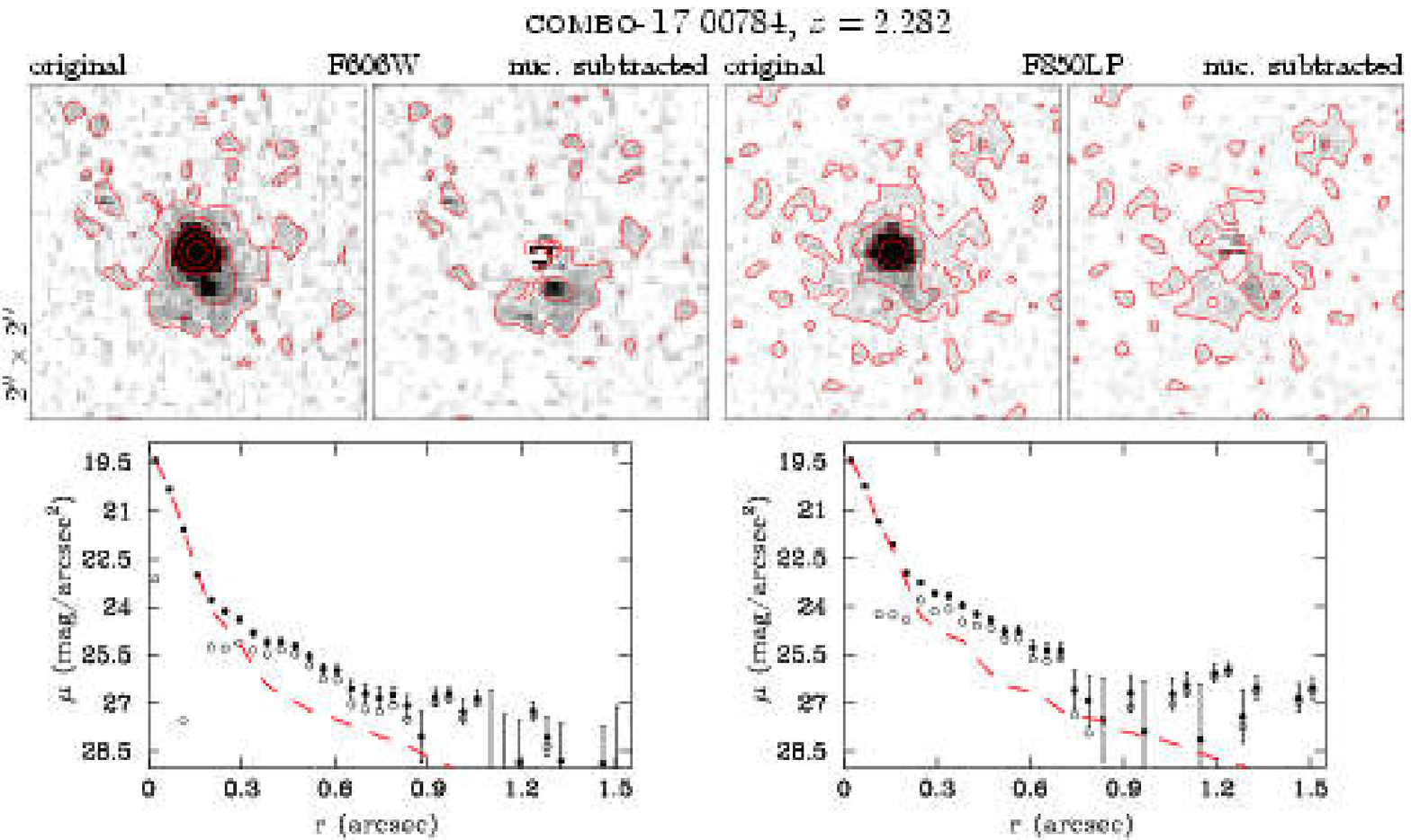}
\addtocounter{figure}{-1}
\caption{
continued
}
\end{figure*}

\begin{figure*}
%
%
\epsscale{1.0}
\plotone{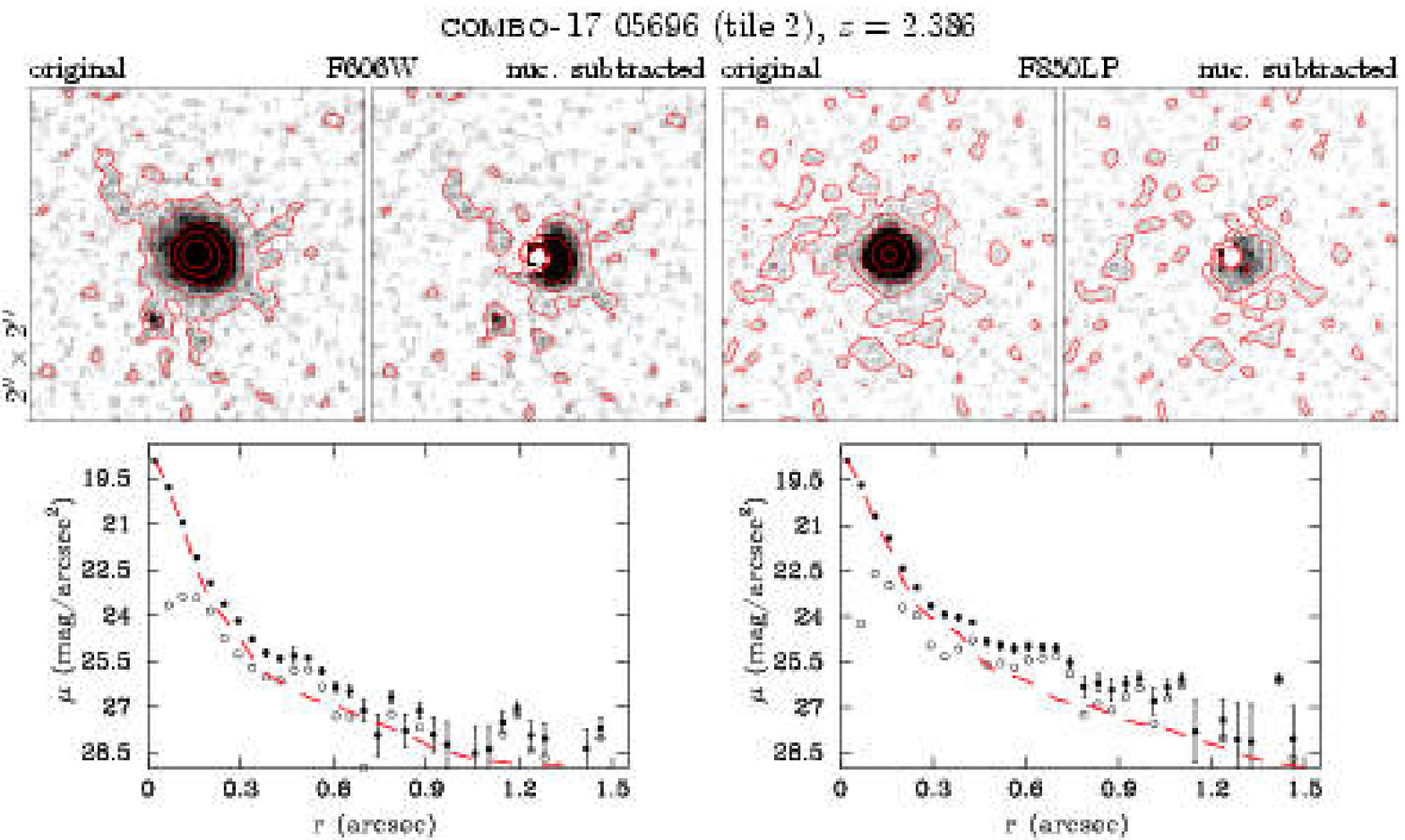}
\plotone{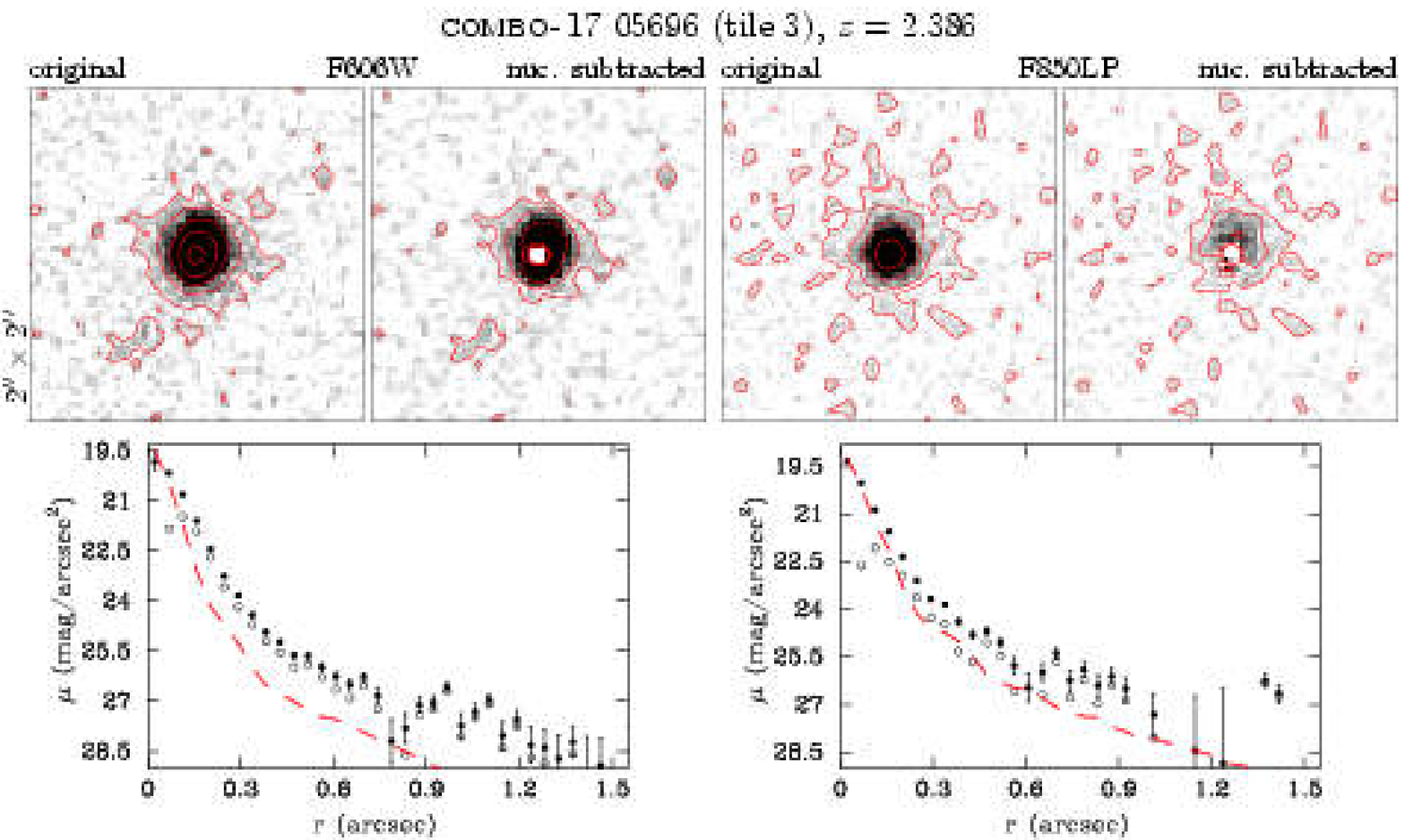}
\addtocounter{figure}{-1}
\caption{
continued
}
\end{figure*}

\begin{figure*}
%
%
\epsscale{1.0}
\plotone{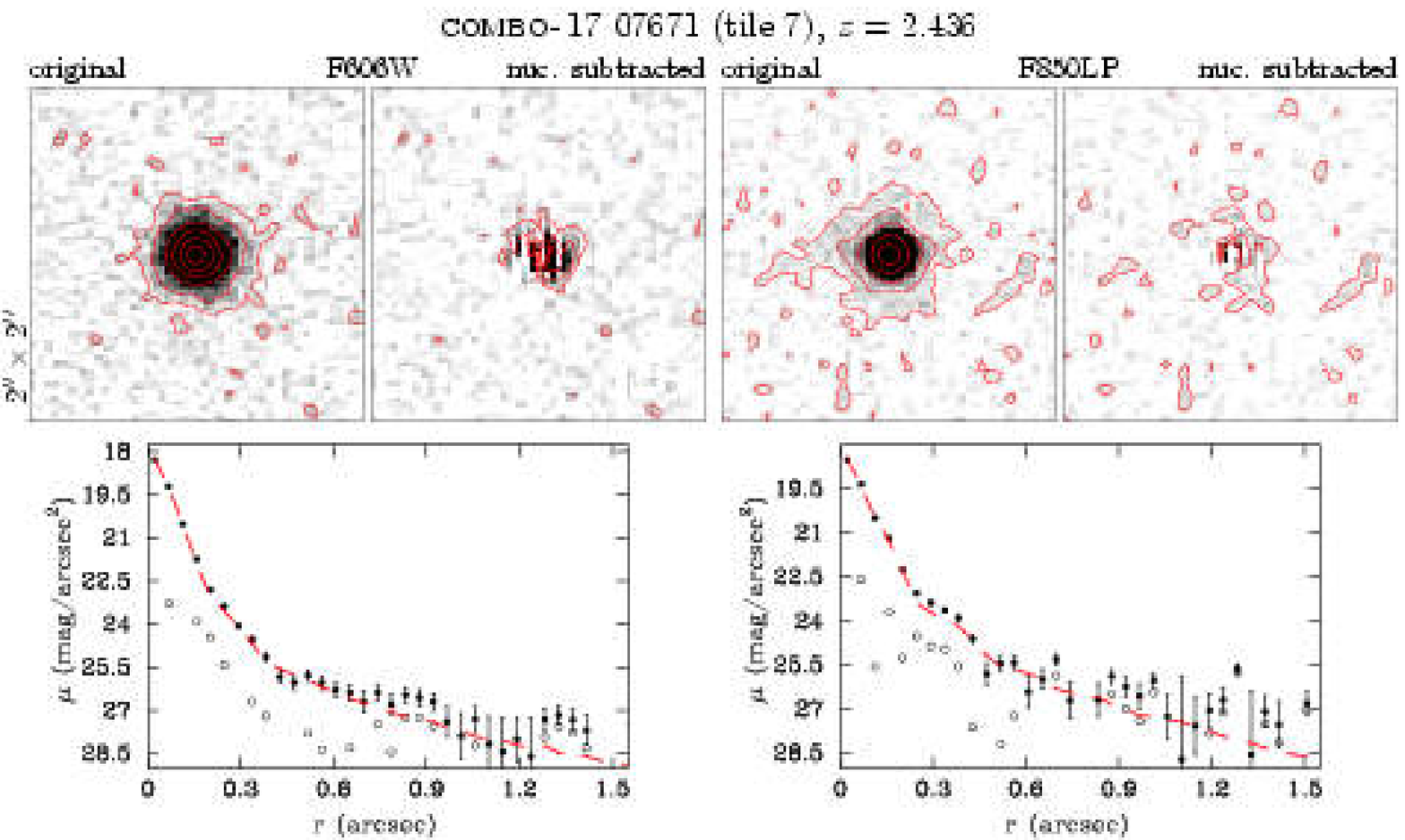}
\plotone{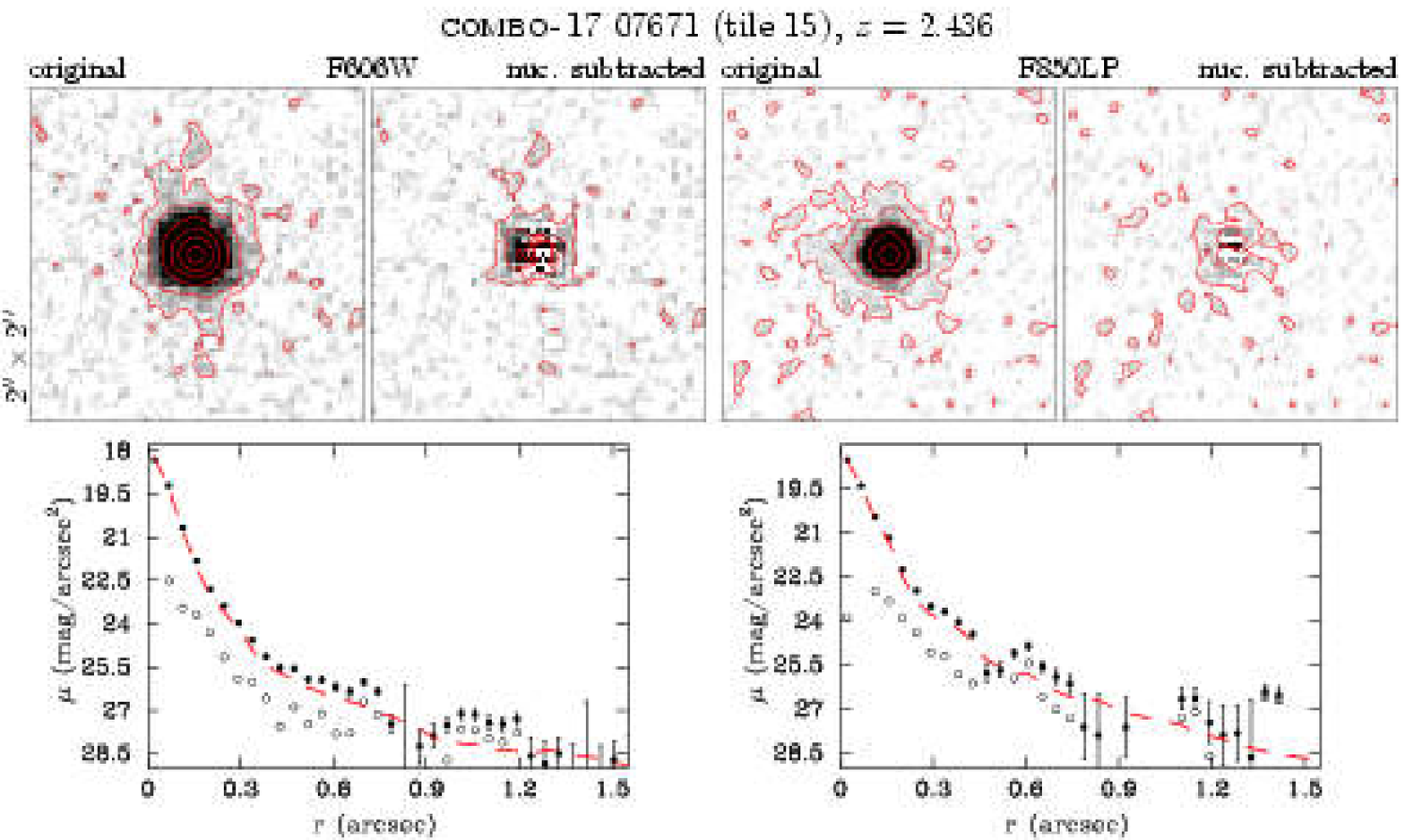}
\addtocounter{figure}{-1}
\caption{
continued
}
\end{figure*}

\begin{figure*}
%
%
\epsscale{1.0}
\plotone{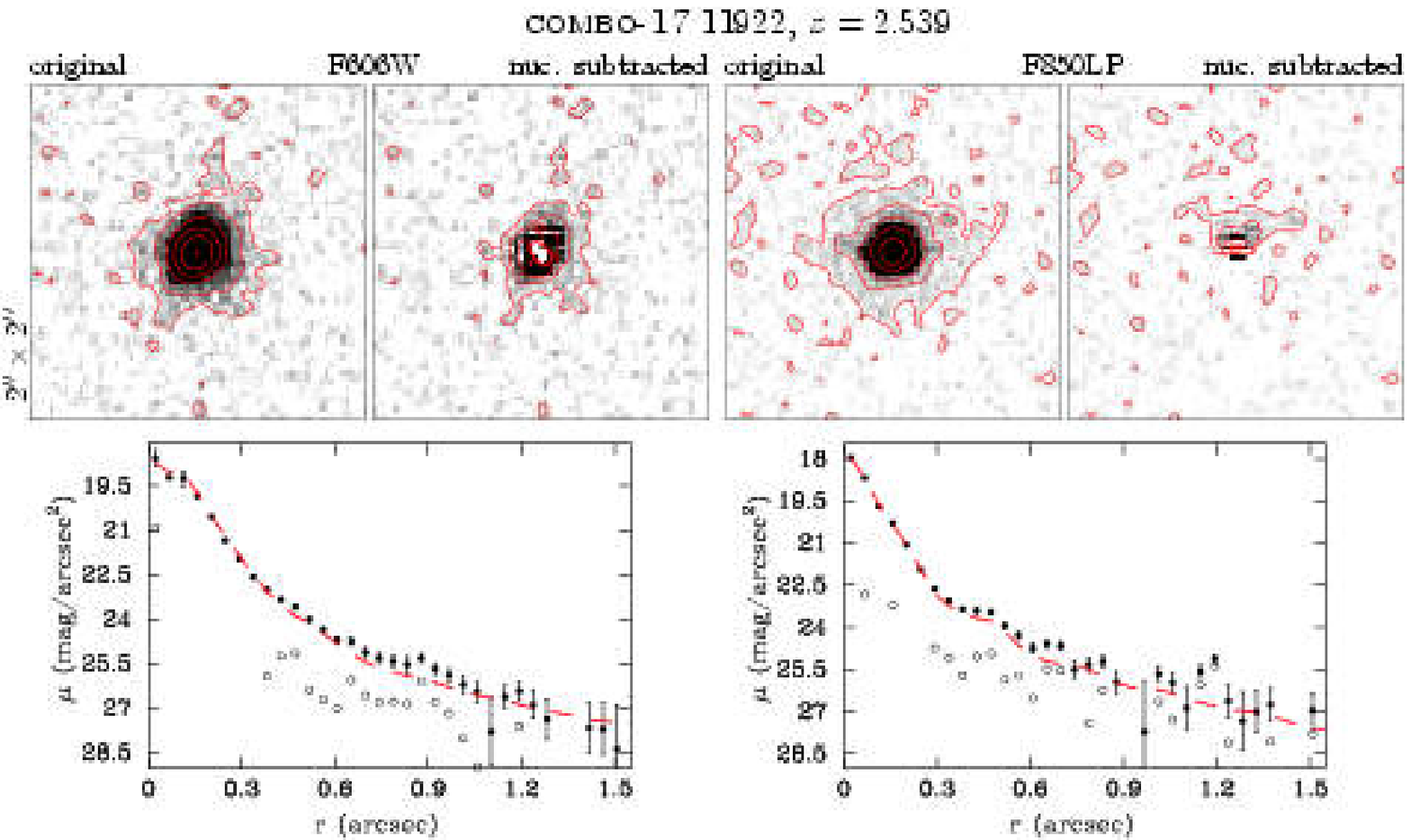}
\plotone{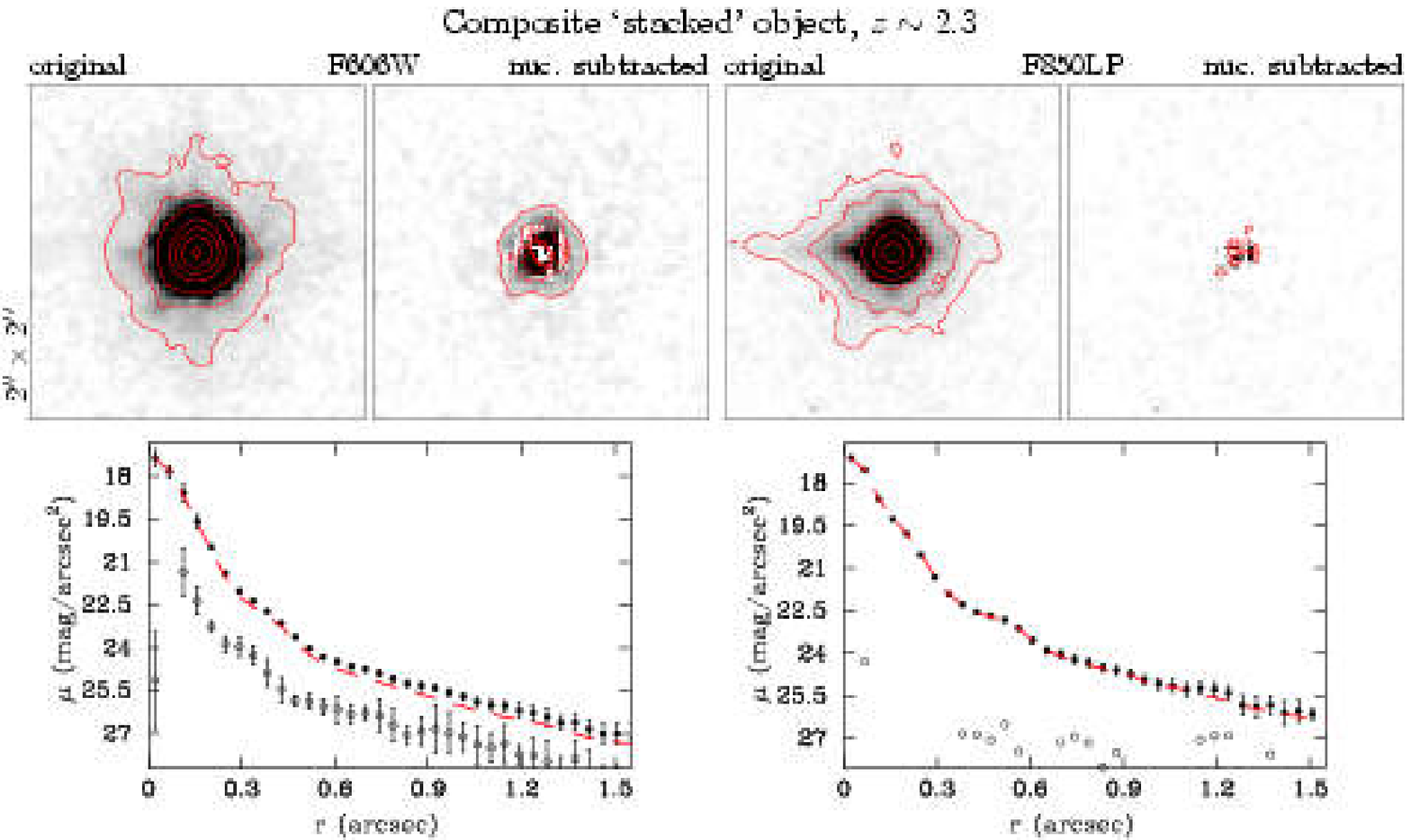}
\addtocounter{figure}{-1}
\caption{
continued
}
\end{figure*}

\begin{figure*}
\includegraphics[clip,width=\fullwidth]{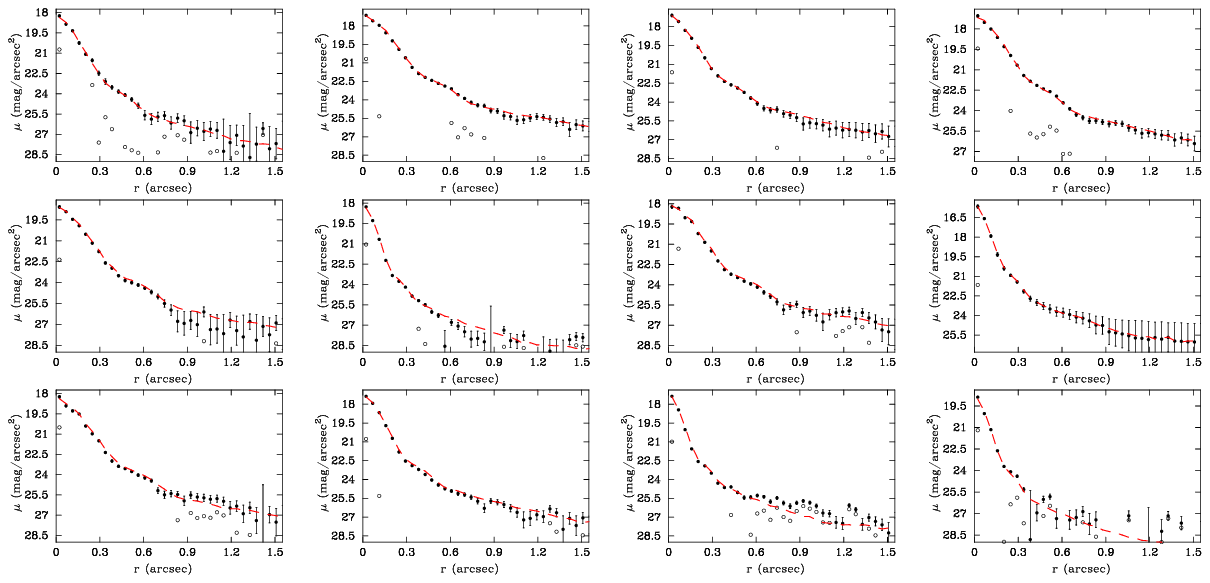}
\caption{\label{fig:stars_v}
PSF peak subtraction applied to field stars: \V-band. Shown are
surface brightness profiles for 12 randomly selected field stars in
the \V-band. Symbols and lines as in Figure~\ref{fig:allimages}.
}
\end{figure*}

\begin{figure*}
\includegraphics[clip,width=\fullwidth]{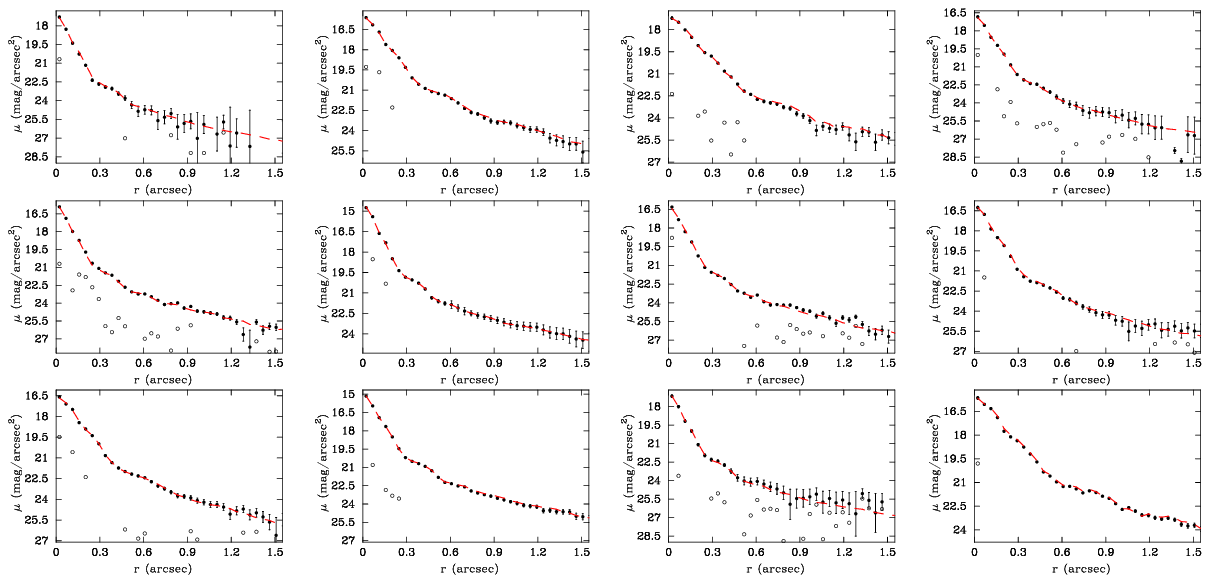}
\caption{\label{fig:stars_z}
As Figure~\ref{fig:stars_v}, but for 12 different randomly selected
stars in the \z-band.
}
\end{figure*}

\end{document}